\documentstyle[aps,prd,eqsecnum]{revtex}

\newcommand{\be}{\begin{equation}}
\newcommand{\ee}{\end{equation}}
\newcommand{\bea}{\begin{eqnarray}}
\newcommand{\eea}{\end{eqnarray}}

\newcommand{\dst}{\displaystyle}
\newcommand{\pa}{\partial}

\def\htt#1#2{h^{\rm TT}_{#1#2}}
\def\httdot#1#2{\dot{h}^{\rm TT}_{#1#2}}

\def\httiv#1#2{h^{\rm TT}_{(4)#1#2}}
\def\httivdot#1#2{\dot{h}^{\rm TT}_{(4)#1#2}}
\def\ghttiv#1#2#3{h^{\rm TT}_{(4)#1#2,#3}}

\def\pitt#1#2{\pi^{#1#2{\rm TT}}}
\def\pitiii#1#2{\widetilde{\pi}^{#1#2}_{(3)}}

\def\np{({\bf n}\cdot{\bf p})}
\def\pp{{\bf p}^2}
\def\ppp{({\bf p}^2)}

\def\omk{\omega_{\text{kinetic}}}
\def\oms{\omega_{\text{static}}}

\begin{document}

\title{Dynamical invariants for general relativistic two-body systems
at the third post-Newtonian approximation}

\author{Thibault Damour}
\address{Institut des Hautes \'Etudes Scientifiques,
91440 Bures-sur-Yvette, France}

\author{Piotr Jaranowski}
\address{Institute of Theoretical Physics,
Bia{\l}ystok University,
Lipowa 41, 15-424 Bia{\l}ystok, Poland}

\author{Gerhard Sch\"afer}
\address{Theoretisch-Physikalisches Institut,
Friedrich-Schiller-Universit\"at,
Max-Wien-Platz 1, 07743 Jena, Germany}

\maketitle

\begin{abstract}
We extract all the invariants (i.e.\ all the functions which do not depend on
the choice of phase-space coordinates) of the dynamics of two point-masses, at
the third post-Newtonian (3PN) approximation of general relativity.  We start by
showing how a contact transformation can be used to reduce the 3PN higher-order
Hamiltonian derived by Jaranowski and Sch\"afer \cite{JS98} to an ordinary
Hamiltonian.  The dynamical invariants for general orbits (considered in the
center-of-mass frame) are then extracted by computing the radial action variable
$\oint{p_r}dr$ as a function of energy and angular momentum.  The important 
case of circular orbits is given special consideration.  We discuss in detail 
the plausible ranges of values of the two quantities $\oms$, $\omk$ which
parametrize the existence of ambiguities in the regularization of some of the
divergent integrals making up the Hamiltonian.  The physical applications of the
invariant functions derived here (e.g.\ to the determination of the location of
the last stable circular orbit) are left to subsequent work.
\end{abstract}


\section{Introduction}

Binary sytems made of compact objects (neutron stars or black holes) are the
most promising candidate sources for ground based interferometric
gravitational-wave detectors such as LIGO and VIRGO.  In the case of (stellar)
black holes the gravitational waveform will enter the detector bandwidth only at
the last stage of the inspiral motion, just before the inspiral turns into a
plunge.  The detection of these gravitational signals will be possible only by
correlating the detector output with sufficiently accurate copies of the 
expected
signals (``matched filtering").  It is therefore very important to have the best
possible analytical control of the dynamics of general relativistic two-body
systems.  The equations of motion of binary systems have been derived some years
ago at the 5/2 post-Newtonian (2.5PN) approximation\footnote{We recall that the
``$n$PN approximation" means the obtention of the terms of order
$(v/c)^{2n}\sim(Gm/c^2r)^n$ in the equations of motion.} \cite{DD}.  Recently,
it has been possible to derive the third post-Newtonian (3PN) dynamics of
point-mass systems \cite{JS98}, though with some remaining ambiguity due to the
need to regularize the divergent integrals caused by the use of
Dirac-delta-function sources.  The knowledge of the conservative part of the
dynamics of binary systems must then be completed by a correspondingly accurate
knowledge of the gravitational-wave luminosity (used, heuristically, to derive
an accurate estimate of the radiation damping effects which drive the inspiral
motion of binary systems).  The gravitational-wave luminosity is currently known
to the {\it fractional} 2.5PN accuracy \cite{BDIWW}.
  By combining the 2PN-level conservative
dynamics \cite{DD} with the 2.5PN gravitational-wave luminosity \cite{BDIWW} one
has recently constructed some (improved) filters (P-approximants) \cite{DIS98},
for application to gravitational-wave data analysis problems.

The main purpose of this work is to extract all the {\it invariants}, i.e., the
functions which do not depend on the choice of coordinates (in space or
phase-space), of the 3PN dynamics derived in Ref.\ \cite{JS98}.  This task is
important for three reasons:  (i) some of the invariant functions are directly
useful for deriving the 3PN ``phasing formula" of inspiralling binaries, i.e.,
for constructing 3PN-accurate gravitational-wave filters; (ii) we shall use, in
a companion paper \cite{DJS2}, some of the invariants derived below to determine
the location of the Last (circular) Stable Orbit which marks the transition
between the inspiral and the plunge, and (iii) from a practical point of view,
the dynamical invariants will be quite useful for comparing the 3PN ADM
Hamiltonian dynamics of \cite{JS98} with forthcoming derivations of the 3PN
equations of motion in harmonic coordinate systems \cite{BFPW}.

\section{Reduction of the 3PN higher-order Hamiltonian}

It was shown some years ago \cite{DS85} that, in most coordinate systems, the
conservative part of the PN-expanded equations of motion of two body systems,
say $\ddot{\bf x}_a={\bbox{\cal A}}_a({\bf x}_b,\dot{\bf x}_b)$, where $a,b=1,2$
and where ${\bbox{\cal A}}
={\bf A}_0+c^{-2}{\bf A}_2+c^{-4}{\bf A}_4+c^{-6}{\bf A}_6+\ldots\,$, do not 
follow from any {\it ordinary} Lagrangian $L({\bf x}_a,\dot{\bf x}_a)$.  For 
instance, in harmonic coordinates, and at the 2PN level, one needs to consider 
an acceleration-dependent Lagrangian $L^{\rm
harmonic}_{\rm 2PN}({\bf x}_a,\dot{\bf x}_a,\ddot{\bf x}_a)$ \cite{DD}.
However, it was shown in Ref.\ \cite{S84}, and, more generally, in Ref.\
\cite{DS85}, that any higher-order PN-expanded Lagrangian $L({\bf x}_a,\dot{\bf
x}_a,\ddot{\bf x}_a,\ldots)$ (where higher derivatives enter only 
perturbatively) can be reduced to an ordinary Lagrangian $L'({\bf
x}'_a,\dot{\bf x}'_a)$ by a suitable (higher-order) {\it contact 
transformation}
${\bf x}'_a(t) ={\bf x}_a(t)-{\bbox{\epsilon}}_a({\bf x}_b,\dot{\bf
x}_b,\ldots)$.  At the 2PN level the class of coordinates where the dynamics
admit an ordinary Lagrangian is rather restricted \cite{DS85}, but it includes
in particular the ADM coordinates \cite{OOKH,S85}.  At the 3PN level, 
Jaranowski
and Sch\"afer, who worked within the ADM canonical formalism, have found that
the (conservative) ADM dynamics could not be derived from an ordinary
Hamiltonian $H({\bf x}_a,{\bf p}_a)$ (equivalent to an ordinary Lagrangian
$L({\bf x}_a,\dot{\bf x}_a)$), but that instead it could be derived from a
certain higher-order Hamiltonian $\widetilde{H}({\bf x}_a,{\bf p}_a,\dot{\bf
x}_a,\dot{\bf p}_a)$.  [This higher-order matter Hamiltonian is defined by
eliminating the field variables ${\htt ij}$, ${\httdot ij}$ in a certain Routh
(i.e., mixed) functional $R({\bf x}_a,{\bf p}_a,{\htt ij},{\httdot ij})$; see
Eq.\ (33) of Ref.\ \cite{JS98}.]  The meaning of this higher-order matter
Hamiltonian $\widetilde{H}({\bf x}_a,{\bf p}_a,\dot{\bf x}_a,\dot{\bf p}_a)$ is 
that
the equations of motion of the matter can be written (after elimination of the
TT variables) as
\be
\label{21}
\dot{\bf x}_a
= \frac{\delta\widetilde{H}({\bf x}_b,{\bf p}_b,\dot{\bf x}_b,\dot{\bf p}_b)} 
{\delta{\bf p}_a},\quad
\dot{\bf p}_a
= -\frac{\delta\widetilde{H}({\bf x}_b,{\bf p}_b,\dot{\bf x}_b,\dot{\bf p}_b)}
{\delta{\bf x}_a},
\ee
where $\delta/\delta{\bf p}_a$ and $\delta/\delta{\bf x}_a$ denote functional 
derivatives:
\be
\label{22}
\frac{\delta\widetilde{H}}{\delta{\bf p}_a} \equiv
\frac{\pa\widetilde{H}}{\pa{\bf p}_a}
- \frac{d}{dt}\left(\frac{\pa\widetilde{H}}{\pa\dot{\bf p}_a}\right),\quad
\frac{\delta\widetilde{H}}{\delta{\bf x}_a} \equiv
\frac{\pa\widetilde{H}}{\pa{\bf x}_a}
- \frac{d}{dt}\left(\frac{\pa\widetilde{H}}{\pa\dot{\bf x}_a}\right).
\ee

It is easily seen that the Hamilton-like equations of motion (\ref{21}) are
equivalent to the Euler-Lagrange equations derived by extremizing the action
functional
\be
\label{23}
\widetilde{S}[{\bf x}_a(t),{\bf p}_a(t)] = \int dt\,
\widetilde{L}({\bf x}_b,{\bf p}_b,\dot{\bf x}_b,\dot{\bf p}_b),
\ee
where
\be
\label{24}
\widetilde{L}({\bf x}_b,{\bf p}_b,\dot{\bf x}_b,\dot{\bf p}_b) \equiv
{\bf p}_a\dot{\bf x}_a - \widetilde{H}({\bf x}_b,{\bf p}_b,\dot{\bf 
x}_b,\dot{\bf
p}_b).
\ee
Indeed, we have
\be
\label{25}
\frac{\delta\widetilde{L}}{\delta{\bf
p}_a} \equiv \dot{\bf x}_a - \frac{\delta\widetilde{H}}{\delta{\bf p}_a},\quad
\frac{\delta\widetilde{L}}{\delta{\bf x}_a} \equiv -\dot{\bf p}_a -
\frac{\delta\widetilde{H}}{\delta{\bf x}_a}.
\ee

To derive the dynamical invariants of the 3PN equations of motion it is
convenient to introduce new coordinates (in phase-space $({\bf x}_a,{\bf
p}_a)$),
\be
\label{26}
{\bf x}'_a = {\bf x}_a + \delta{\bf x}_a,\quad
{\bf p}'_a = {\bf p}_a + \delta{\bf p}_a,
\ee
such that the equations of motion for $({\bf x}'_a,{\bf p}'_a)$ become some 
ordinary Hamilton equations
\be 
\label{27}
\dot{\bf x}'_a = \frac{\pa H'({\bf x}'_b,{\bf p}'_b)}{\pa{\bf p}'_a},\quad
\dot{\bf p}'_a = -\frac{\pa H'({\bf x}'_b,{\bf p}'_b)}{\pa{\bf x}'_a}.
\ee
This is
equivalent to requiring that the $({\bf x}'_a,{\bf p}'_a)$ equations of motion
derive from an action functional of the form
\be
\label{28}
S'[{\bf x}'_a(t),{\bf p}'_a(t)] = \int dt\,
L'({\bf x}'_b,{\bf p'}_b,\dot{\bf x}'_b,\dot{\bf p}'_b),
\ee
where\footnote{As defined $L'$ does not depend on $\dot{\bf p}'$, but, by 
symmetry, it is convenient to allow for such a dependence (which can be easily 
introduced by transforming ${\bf p}'\dot{\bf x}'$ by parts).}
\be
\label{29}
L'({\bf x}'_b,{\bf p}'_b,\dot{\bf x}'_b,\dot{\bf p}'_b)
\equiv {\bf p}'_a\dot{\bf x}'_a - H'({\bf x}'_b,{\bf p}'_b).
\ee
When so expressed at the level of action functionals, the problem of reducing 
the `higher-order' action $\widetilde{S}[{\bf x}_a(t),{\bf p}_a(t)]$ to an 
ordinary 
action $S'[{\bf x}_a'(t),{\bf p}_a'(t)]$ is quite similar to the problem of the 
order-reduction of (perturbative) higher-order actions
$S[{\bf x}_a(t)]=\int dt\,L({\bf x}_a,\dot{\bf x}_a,\ddot{\bf x}_a,\ldots)$ 
which was solved in full generality in Ref.\ \cite{DS85}.  It is then an easy 
task to adapt the techniques used in \cite{DS85} to solve our present problem.  

First, we note that, taking into account the fact that the Hamiltonian has the 
perturbative structure,
$\widetilde{H}({\bf x}_a,{\bf p}_a,\dot{\bf x}_a,\dot{\bf p}_a)
=H_0({\bf x}_a,{\bf p}_a)
+c^{-2}H_2({\bf x}_a,{\bf p}_a)
+c^{-4}H_4({\bf x}_a,{\bf p}_a)
+c^{-6}H_6({\bf x}_a,{\bf p}_a,\dot{\bf x}_a,\dot{\bf p}_a)$,
the identities (\ref{25}) for the variational derivatives of
$\widetilde{L}({\bf x}_a,{\bf p}_a,\dot{\bf x}_a,\dot{\bf p}_a)$ imply
\begin{mathletters}
\label{210}
\bea
\dot{\bf x}_a \equiv {\bf v}_{{\rm
2PN}a}({\bf x},{\bf p}) + \frac{\delta\widetilde{L}}{\delta{\bf p}_a} + {\cal
O}(c^{-6}) = {\bf v}_{{\rm N}a}({\bf x},{\bf p}) +
\frac{\delta\widetilde{L}}{\delta{\bf p}_a} + {\cal O}(c^{-2}),
\\[2ex]
\dot{\bf
p}_a \equiv {\bf q}_{{\rm 2PN}a}({\bf x},{\bf p}) -
\frac{\delta\widetilde{L}}{\delta{\bf x}_a} + {\cal O}(c^{-6}) = {\bf q}_{{\rm
N}a}({\bf x},{\bf p}) - \frac{\delta\widetilde{L}}{\delta{\bf x}_a} + {\cal
O}(c^{-2}).
\eea
\end{mathletters}
Here ${\bf v}_{ia}$ and ${\bf q}_{ia}$ (with
$i$ = 2PN or N) denote some explicit functions of ${\bf x}$ and ${\bf p}$ which
are the right-hand-sides of the usual Hamilton equations at the 2PN (or
Newtonian) level.  For instance, at the Newtonian level ${\bf v}_{{\rm 
N}a}({\bf
x},{\bf p})={\bf p}_a/m_a$, ${\bf q}_{{\rm N}a}({\bf x},{\bf p}) =-Gm_am_b({\bf
x}_a-{\bf x}_b)/|{\bf x}_a-{\bf x}_b|^3$.  Note that in Eqs.\ (\ref{210}), and
in the reasoning below, we are working `off shell', i.e., we consider virtual
motions which do not necessarily satisfy the equations of motion. 
 [As we are using
identities for action functionals, it is essential to work off shell.]
Inserting the perturbative identities (\ref{210}) in $\widetilde{H}({\bf x},{\bf
p},\dot{\bf x},\dot{\bf p})$ and Taylor expanding yields the
identity\footnote{In Eq.\ (\ref{211}) and elsewhere, we abbreviate the notation
by suppressing the (summed over) indices in $\dot{\bf x}=\dot{x}_a^i$, ${\bf
p}=p_{ai}$, etc.}
\be
\label{211}
\widetilde{H}({\bf x},{\bf p},\dot{\bf x},\dot{\bf p})
\equiv \bar{H}({\bf x},{\bf p}) +
\frac{\pa\widetilde{H}}{\pa\dot{\bf x}} \frac{\delta\widetilde{L}}{\delta{\bf 
p}} -
\frac{\pa\widetilde{H}}{\pa\dot{\bf p}} \frac{\delta\widetilde{L}}{\delta{\bf 
x}} +
\,\mbox{double-zeros}\, + {\cal O}(c^{-8}),
\ee
where $\bar{H}({\bf x},{\bf p})$
denotes the naive `order-reduced' Hamiltonian obtained by using the
(lower-order) equations of motion to eliminate the higher-order derivative 
terms
(a wrong procedure in general):
\be \label{212} \bar{H}({\bf x},{\bf p}) 
\equiv
\widetilde{H}({\bf x},{\bf p}, {\bf v}_{{\rm N}}({\bf x},{\bf p}),{\bf q}_{{\rm
N}}({\bf x},{\bf p})) + {\cal O}(c^{-8}).  \ee [As indicated in Eq.\
(\ref{212}), it is sufficient to use the Newtonian equations of motion because
$\dot{\bf x}$ and $\dot{\bf p}$ enter only at the $c^{-6}$ level.]  The
`double-zeros' in Eq.\ (\ref{211}) denote all the terms generated by the Taylor
expansion which would be at least quadratic in $\delta\widetilde{L}/\delta{\bf 
x}$
and $\delta\widetilde{L}/\delta{\bf p}$.  As is well known \cite{BOC80,DS85} 
such
terms can be perturbatively neglected (even off shell) because they do not
contribute to the equations of motion at the 3PN level.

If one inserts the identity (\ref{211}) in the original action functional
(\ref{23}), one sees that $\widetilde{S}[{\bf x},{\bf p}]$ has the desired 
ordinary
form (\ref{28}) modulo some extra terms which are {\it linear} in
$\delta\widetilde{L}/\delta{\bf x}$ and $\delta\widetilde{L}/\delta{\bf p}$.  As 
in
Ref.\ \cite{DS85} such terms can be eliminated by noticing that (to first order)
 the effect of
the shift of dynamical variables (\ref{26}) on the Lagrangian 
$\widetilde{L}({\bf
x},{\bf p},\dot{\bf x},\dot{\bf p})$ reads (by virtue of the definition of
functional derivatives) \be \label{213} \widetilde{L}({\bf x},{\bf p},\dot{\bf
x},\dot{\bf p}) \equiv \widetilde{L}({\bf x}',{\bf p}',\dot{\bf x}',\dot{\bf 
p}') -
\frac{\delta\widetilde{L}}{\delta{\bf x}}\delta{\bf x} -
\frac{\delta\widetilde{L}}{\delta{\bf p}}\delta{\bf p} + 
\frac{d}{dt}Q(\delta{\bf
x},\delta{\bf p}), \ee where $Q(\delta{\bf x},\delta{\bf p})$ is some linear
form in $\delta{\bf x}$ and $\delta{\bf p}$. [It is sufficient to work to 
linear order because $\delta{\bf x}$ and $\delta{\bf p}$ are 
${\cal O}(c^{-6})$.]

  By combining the two identities
(\ref{211}) and (\ref{213}) we find that, if we define the ordinary Lagrangian
(considered in phase-space) $\bar{L}$, associated to the ordinary (naively
reduced) Hamiltonian $\bar{H}({\bf x},{\bf p})$, Eq.\ (\ref{212}), \be
\label{214} \bar{L}({\bf x},{\bf p},\dot{\bf x},\dot{\bf p}) \equiv {\bf p}_a
\dot{\bf x}_a - \bar{H}({\bf x},{\bf p}), \ee we have the identity \be
\label{215} \widetilde{L}({\bf x},{\bf p},\dot{\bf x},\dot{\bf p}) \equiv
\bar{L}({\bf x}',{\bf p}',\dot{\bf x}',\dot{\bf p}') -
\frac{\delta\widetilde{L}}{\delta{\bf x}} \left(\delta{\bf
x}-\frac{\pa\widetilde{H}}{\pa\dot{\bf p}}\right) -
\frac{\delta\widetilde{L}}{\delta{\bf p}} \left(\delta{\bf
p}+\frac{\pa\widetilde{H}}{\pa\dot{\bf x}}\right) + \frac{d}{dt}Q(\delta{\bf
x},\delta{\bf p}) + \,\mbox{double-zeros}\, + {\cal O}(c^{-8}).
\ee
Therefore, if we shift the phase-space coordinates by defining
\be
\label{216}
{\bf x}'_a-{\bf x}_a \equiv \delta{\bf x}_a
= \frac{\pa\widetilde{H}}{\pa\dot{\bf p}_a},\quad
{\bf p}'_a - {\bf p}_a \equiv \delta{\bf p}_a
= -\frac{\pa\widetilde{H}}{\pa\dot{\bf x}_a},
\ee
we find (noticing that both total differentials, and double-zeros, are
negligible in action functionals) that the original equations of motion
(\ref{21}) are transformed, when rewritten in $({\bf x}',{\bf p}')$ 
coordinates,
into the Euler-Lagrange equations of the `ordinary' phase-space Lagrangian
$\bar{L}({\bf x}',{\bf p}',\dot{\bf x}',\dot{\bf p}')$, i.e.\ into ordinary
Hamilton equations \be \label{217} \dot{\bf x}' = \frac{\pa\bar{H}({\bf 
x}',{\bf
p}')}{\pa{\bf p}'} + {\cal O}(c^{-8}),\quad \dot{\bf p}' =
-\frac{\pa\bar{H}({\bf x}',{\bf p}')}{\pa{\bf x}'} + {\cal O}(c^{-8}).  \ee

Summarizing:  it is licit to naively reduce the higher-order original
Hamiltonian $\widetilde{H}({\bf x},{\bf p},\dot{\bf x},\dot{\bf p})$ by 
replacing
the lower-order equations of motion in the offending derivatives $\dot{\bf x}$,
$\dot{\bf p}$ (thereby defining the reduced Hamiltonian $\bar{H}({\bf x},{\bf
p})$, Eq.\ (\ref{212})), if one adds the correcting information that the
ordinary Hamilton equations defined by the reduced $\bar{H}({\bf x},{\bf p})$
hold in the new phase-space coordinates $({\bf x}',{\bf p}')$ defined by Eq.\
(\ref{216}).

Let us note for completness that the conserved energy ${\cal E}'=\bar{H}({\bf
x}',{\bf p}')$ defined by the autonomous Hamiltonian $\bar{H}$ becomes, when
expressed in the original variables \be \label{218} {\cal E}'=\bar{H}({\bf
x}',{\bf p}')=\bar{H}({\bf x},{\bf p}) +\frac{\pa\bar{H}}{\pa{\bf
x}}\frac{\pa\widetilde{H}}{\pa\dot{\bf p}} -\frac{\pa\bar{H}}{\pa{\bf
p}}\frac{\pa\widetilde{H}}{\pa\dot{\bf x}} ={\cal E}+{\cal O}(c^{-8}), \ee where 
\be
\label{219} {\cal E}=\widetilde{H}({\bf x},{\bf p},\dot{\bf x},\dot{\bf p})
-\dot{\bf x}\frac{\pa\widetilde{H}}{\pa\dot{\bf x}} -\dot{\bf
p}\frac{\pa\widetilde{H}}{\pa\dot{\bf p}} \ee is, indeed, easily checked to be a
conserved quantity associated with the higher-order dynamics (\ref{21}).

Let us now consider the explicit application of the previous results to the 
case
at hand.  Following \cite{JS98} it is sufficient to consider the dynamics of 
the
relative motion of a two-body system, considered in the center-of-mass frame
(${\bf p}_1+{\bf p}_2=0$).  It is convenient to work with the following reduced
variables\footnote{Note that \cite{JS98} use units where $16\pi G=1$.}  \be
\label{220} {\bf r} \equiv \frac{{\bf x}_1-{\bf x}_2}{GM},\quad {\bf p} \equiv
\frac{{\bf p}_1}{\mu} = -\frac{{\bf p}_2}{\mu},\quad \widehat{t} \equiv
\frac{t}{GM},\quad \widehat{H}^{\rm NR} \equiv \frac{\widetilde{H}^{\rm 
NR}}{\mu}, \ee
where \be \label{221} M \equiv m_1 + m_2,\quad \mu \equiv 
\frac{m_1m_2}{M},\quad
\nu \equiv \frac{\mu}{M} = \frac{m_1m_2}{(m_1+m_2)^2}.  \ee In Eq.\ (\ref{220})
the superscript NR denotes a `non-relativistic' (higher-order) Hamiltonian,
i.e.\ the Hamiltonian without the rest-mass contribution $Mc^2$.  
$\widehat{H}^{\rm
NR}$ is, to start with, a function of ${\bf r}$, ${\bf p}$, $\dot{\bf r}$, 
and
$\dot{\bf p}$.  Here the dot denotes the reduced-time derivative:  \be
\label{222} \dot{\bf r} = \frac{d{\bf r}}{d\widehat{t}} = \frac{d({\bf x}_1-{\bf
x}_2)}{dt},\quad \dot{\bf p} = \frac{d{\bf p}}{d\widehat{t}} =
\frac{G}{\nu}\frac{d{\bf p}_1}{dt} = -\frac{G}{\nu}\frac{d{\bf p}_2}{dt}.  \ee

{}From Eq.\ (\ref{216}) above and Eq.\ (68) of \cite{JS98} one finds that the
coordinate shift needed to transform $\widehat{H}^{\rm NR}({\bf r},{\bf 
p},\dot{\bf
r},\dot{\bf p})$ into an ordinary Hamiltonian reads (on shell\footnote{Working
on shell here is equivalent to neglecting some double-zero terms, see, e.g.\
Ref.\ \cite{DS85}.}, i.e.\ by using the (Newtonian) equations of motion after
the differentiations exhibited in Eq.\ (\ref{216}))
\begin{mathletters}
\label{223}
\bea
\delta{\bf r} &\equiv& {\bf r}'-{\bf r}
= \left.\frac{\pa\widehat{H}^{\rm NR}}{\pa\dot{\bf p}}\right|_{\text{on shell}}
= \frac{1}{c^6} \left\{
\left[ \frac{1}{8}\nu^2\biglb(\np^4-\ppp^2\bigrb)
+\frac{1}{24}\biglb((19-26\nu)\nu\,\np^2-5(13+4\nu)\nu\,\pp\bigrb)\frac{1}{r} 
\right]{\bf n}
\right.\nonumber\\[2ex]&&\left.
+ \frac{1}{12} \left[ \nu^2\biglb(3\pp+\np^2\bigrb)\np
-(14-\nu)\nu\,\np\frac{1}{r} \right]{\bf p} \right\},
\\[2ex]
\delta{\bf p} &\equiv& {\bf p}'-{\bf p}
= -\left.\frac{\pa\widehat{H}^{\rm NR}}{\pa\dot{\bf r}}\right|_{\text{on shell}}
= \frac{1}{c^6} \left\{ \left[ 
\frac{1}{8}\nu^3\biglb(\ppp^2+2\np^2\pp+5\np^4\bigrb)\np\frac{1}{r}
\right.\right.\nonumber\\[2ex]&&\left.
-\frac{1}{48}\biglb(3(1-9\nu)\nu\,\pp
+5(17-29\nu)\nu\,\np^2\bigrb)\np\frac{1}{r^2}
+\frac{1}{512}\biglb(11408-915\pi^2+36(48-5\pi^2)\nu\bigrb)\nu\,\np\frac{1}{r^3}
\right]{\bf n}
\nonumber\\[2ex]&& + \left[
-\frac{1}{8}\nu^3\biglb(\ppp^2+2\np^2\pp+5\np^4\bigrb)\frac{1}{r}
+\frac{1}{16}\biglb((4-15\nu)\nu\,\pp+(6+7\nu)\nu\,\np^2\bigrb)\frac{1}{r^2}
\right.\nonumber\\[2ex]&&\left.\left.
+\frac{1}{1536}\biglb(15(61\pi^2-880)+4(45\pi^2-592)\nu\bigrb)\nu\frac{1}{r^3}
\right]{\bf p} \right\}.
\eea
\end{mathletters}
Note again that $\delta{\bf r}$ and $\delta{\bf p}$ are of 3PN order.

\section{Order-reduced 3PN Hamiltonian}

As explained above the phase-space coordinate transformation (\ref{223}) maps
the original higher-order 3PN ADM dynamics onto ordinary Hamilton equations, 
with an Hamiltonian $\widehat{\bar{H}}({\bf r}',{\bf p}')$ defined by the 
`naive' procedure (\ref{212}), i.e.\ by using in
$\widehat{H}^{\rm NR}({\bf r},{\bf p},\dot{\bf r},\dot{\bf p})$ the following 
replacement rules (in our reduced units where
$\widehat{H}^{\rm NR}_{\text{Newtonian}}({\bf r},{\bf p})={\bf p}^2/2-1/r$, with
$r\equiv|{\bf r}|$)
\be
\label{31}
{\bf r} \to {\bf r}',\quad
{\bf p} \to {\bf p}',\quad
\dot{\bf r} \to {\bf p}',\quad
\dot{\bf p} \to -\frac{{\bf r}'}{r'^{3}}.
\ee
To simplify the writing we shall henceforth drop all the
primes (but the reader should remember that we henceforth work in the shifted
coordinate system $({\bf r}',{\bf p}')$).  We also drop the overbar, on the
Hamiltonian, indicating that we have used the replacement rules (\ref{31}).
Finally, using Eqs.\ (68) and (71) of \cite{JS98}, we obtain the following
ordinary Hamiltonian (with ${\bf n}\equiv{\bf r}/r$)
\be
\label{32}
\widehat{H}^{\text{NR}}\left({\bf r},{\bf p}\right) = \widehat{H}_{\rm 
N}\left({\bf
r},{\bf p}\right) + \frac{1}{c^2} \widehat{H}_{\rm 1PN}\left({\bf r},{\bf 
p}\right)
+ \frac{1}{c^4} \widehat{H}_{\rm 2PN}\left({\bf r},{\bf p}\right) + 
\frac{1}{c^6}
\widehat{H}_{\rm 3PN}\left({\bf r},{\bf p}\right),
\ee
where
\bea
&\label{33}\dst
\widehat{H}_{\rm N}\left({\bf r},{\bf p}\right) = \frac{\pp}{2} - \frac{1}{r},
&\\[2ex]&\dst
\label{34}
\widehat{H}_{\rm 1PN}\left({\bf r},{\bf p}\right) =
\frac{1}{8}(3\nu-1) \ppp^2 - \frac{1}{2}\left[(3+\nu)\pp + \nu \np^2
\right]\frac{1}{r} + \frac{1}{2r^2},
&\\[2ex]&\dst
\label{35}
\widehat{H}_{\rm 2PN}\left({\bf r},{\bf p}\right) = 
\frac{1}{16}\left(1-5\nu+5\nu^2\right) 
\ppp^3
+ \frac{1}{8}\left[ \left(5-20\nu-3\nu^2\right)\ppp^2 - 2\nu^2 \np^2\pp - 
3\nu^2
\np^4 \right]\frac{1}{r}
\nonumber&\\[2ex]&\dst
+ \frac{1}{2} \left[(5+8\nu)\pp+3\nu\np^2\right]\frac{1}{r^2}
- \frac{1}{4}(1+3\nu)\frac{1}{r^3},
&\\[2ex]&\dst
\label{36}
\widehat{H}_{\rm 3PN}\left({\bf r},{\bf p}\right)
= \frac{1}{128}\left(-5+35\nu-70\nu^2+35\nu^3\right)\ppp^4
\nonumber&\\[2ex]&\dst
+ \frac{1}{16}\left[
\left(-7+42\nu-53\nu^2-5\nu^3\right)\ppp^3
+ (2-3\nu)\nu^2\np^2\ppp^2
+ 3(1-\nu)\nu^2\np^4\pp - 5\nu^3\np^6
\right]\frac{1}{r}
\nonumber&\\[2ex]&\dst
+\left[ \frac{1}{16}\left(-27+136\nu+109\nu^2\right)\ppp^2
+ \frac{1}{16}(17+30\nu)\nu\np^2\pp + \frac{1}{12}(5+43\nu)\nu\np^4
\right]\frac{1}{r^2}
&\nonumber\\[2ex]&\dst
+\left\{ \left[ 
-\frac{25}{8}+(\frac{1}{64}\pi^2-\frac{335}{48})\nu-\frac{55}{12}\nu^2
\right]\pp
+ \left(-\frac{85}{16}-\frac{3}{64}\pi^2+\frac{27}{8}\nu\right)\nu\np^2 
\right\}\frac{1}{r^3}
&\nonumber\\[2ex]&\dst
+ \left[\frac{1}{8}+(\frac{109}{12}-\frac{21}{32}\pi^2)\nu\right]\frac{1}{r^4}
+ \omk\left(\pp-3\np^2\right)\frac{\nu^2}{r^3} + \oms \frac{\nu}{r^4}.
\eea
The parameters $\omk$ and $\oms$ appearing in the 3PN Hamiltonian
$\widehat{H}_{\rm 3PN}$ parametrize the existence of {\it ambiguities} in the 
regularization (as it is presently performed) of some of the divergent integrals 
making up the Hamiltonian.

It was shown by Damour \cite{leshouches} that the internal structure of the
compact bodies making up a binary system (e.g.\ a neutron star versus a black
hole) start influencing the dynamics only at the 5PN level ($G^6/c^{10}$; see
Eq.\ (19) in Sec.\ 5 of Ref.\ \cite{leshouches}).  Therefore, one expects that
the present 3PN ambiguities are of technical nature, and not linked to real
physical ambiguities.  The best way to resolve these ambiguities would be to
work out in detail, at the 3PN level, the general matching method outlined in
Ref.\ \cite{leshouches}, and applied there at the 2PN level.  However, the
technical difficultities of working at the 3PN level are such that this route
has not yet been attempted.

The `kinetic ambiguity' $\omk\nu^2\left(\pp-3\np^2\right)/r^3$ (proportional to 
the kinetic energy $\pp$) was explicitly introduced in Ref.\ \cite{JS98} (see 
Eq.\ (69) there; with $\omega^{\text{there}}$ corresponding to 
$\omk^{\text{here}}$ modulo the shift of normalization discussed below).  The 
`static ambiguity' $\oms\nu/r^4$ was discussed in Ref.\ \cite{JS99}.  In the 
present paper, we normalize the static ambiguity such that the value $\oms=0$
corresponds to the static interaction\footnote{\label{ft:static}By `static 
interaction' we mean the part of the Hamiltonian which remains when setting to 
zero both the momenta ${\bf p}_1$, ${\bf p}_2$ {\it and} the independent 
gravitational-wave degrees of freedom ${\htt ij}$, ${\pitt ij}$ (time-symmetric 
conformally flat data).}
between two `Brill-Lindquist black holes' (considered instantaneously at rest;
see \cite{JS99}).  The normalization of the kinetic ambiguity used in Eq.\ 
(\ref{36}) of this paper also differs from that used in Eq.\ (71) of 
\cite{JS98}. 
The motivation for this new normalization is given in Appendix A of the present 
paper, where we calculate, following the route of Ref.\ \cite{JS98}, a new 
`reference' Hamiltonian which corresponds to this normalization. In terms of 
these two normalizations, the Hamiltonian explicitly written in \cite{JS98} 
corresponds to the values $\oms=1/8$ and $\omk=7/16 + \omega$; this means that
(after having applied the replacement rule (\ref{31}) and dropped the primes)
\be
\widehat{H}_{\rm 3PN}^{\text{this paper}}\left({\bf r},{\bf p}\right)
= \widehat{H}_{\rm 3PN}^{\text{Ref.\ \cite{JS98}}}\left({\bf r},{\bf p}\right)
+ \left(\oms-\frac{1}{8}\right)\frac{\nu}{r^4}
+ \left(\omk-\frac{7}{16} -\omega\right)
\left(\pp-3\np^2\right)\frac{\nu^2}{r^3}. 
\ee

Though the two parameters $\oms$ and $\omk$ appear separately in Eq.\ 
(\ref{36}), all the dynamical invariants of the 3PN Hamiltonian involve only the 
combination
\be
\label{37}
\sigma(\nu) \equiv \nu\oms+\nu^2\omk.
\ee
Our results below will verify explicitly that only the combination $\sigma$
appears, but it is, in fact, easy to understand why.  Indeed, as we shall
calculate the dynamical invariants of $H({\bf r},{\bf p})$, our results would 
be
the same if we transformed $H$ by an arbitrary canonical transformation.  If we
consider `a small' canonical transformation of order $1/c^6$, it is enough to
work at linear order.  At linear order a canonical transformation is
parametrized by an arbitrary generating function $g({\bf r},{\bf p})$, and its
effect on the Hamiltonian is to change it by the Poisson bracket $\{g,H\}$:  
\be
\label{38}
\delta H({\bf r},{\bf p}) = \{g,H\} = \frac{\pa g}{\pa x^i}\frac{\pa
H}{\pa p_i} - \frac{\pa g}{\pa p_i}\frac{\pa H}{\pa x^i} = p_i \frac{\pa g}{\pa
x^i} - \frac{x^i}{r^3}\frac{\pa g}{\pa p_i} + {\cal O}(g/c^2).
\ee
If we choose $g({\bf r},{\bf p})=\lambda{\bf r}\cdot{\bf p}/r^3$ we find
\be
\label{39}
\delta H = \frac{\lambda}{r^3}\left(\pp-3\np^2\right) - \frac{\lambda}{r^4}.
\ee 
Therefore by choosing (for instance) $\lambda=-\nu^2\omk$, we see that we can 
eliminate $\omk$ at the cost of replacing $\nu\oms$ by 
$\nu\oms-\lambda=\nu\oms+\nu^2\omk$.

An important issue, when discussing the consequences of the 3PN Hamiltonian
(\ref{32}), is the influence of the lack of precise knowledge of the combination
$\sigma$, Eq.\ (\ref{37}), on physical observables.  First, let us emphasize
that, among the hundreds of contributions to the 3PN Hamiltonian which have been
computed in Ref.\ \cite{JS98}, most of them, though they are in general given by
a divergent integral, seem to be regularizable in an unambigous manner.  Indeed,
if one considers {\it separately} the divergences near one particle (say in a
volume $V_1$ near particle 1), and regularizes them `\`a la Riesz', i.e.\ by
introducing a regularizing factor
$(|{\bf x}-{\bf x}_1|/l_1)^{\epsilon_1}=(r_1/l_1)^{\epsilon_1}$ in the 
integrand, the analytic continuation in $\epsilon_1$ of the regularized integral 
$I_1(\epsilon_1)
=\int_{V_1}d^3x\,(r_1/l_1)^{\epsilon_1}F({\bf x},{\bf x}_1,{\bf x}_2)$ does not
contain a pole $\propto1/\epsilon_1$ for most terms.  This suggests that the
well-defined analytic continuation of $I_1(\epsilon_1)$ down to $\epsilon_1=0$
defines, without apparent ambiguity, the regularized value of
$\int_{V_1}d^3x\,F({\bf x},{\bf x}_1,{\bf x}_2)$.  [In the notation of Appendix
B of \cite{JS98} we are considering that $\nu^{\rm JS}=0$ (i.e.\
$\epsilon_2=\nu^{\rm JS}\epsilon=0$), in a $V_1$-restricted integral which
behaves as $A_1+C_1\nu^{\rm JS}/\mu^{\rm JS}=A_1+C_1\epsilon_2/\epsilon_1$,
before taking the limit $\epsilon_1=\mu^{\rm JS}\epsilon\to0$.]

However, a limited class of `dangereous' integrals involve a simple pole as
$\epsilon_1\to0$:
$I_1(\epsilon_1)=Z_1\biglb(\epsilon_1^{-1}+\ln(R_1/l_1)\bigrb)+A_1$ (where the
`infrared' lengthscale $R_1$ is linked to the size of the volume $V_1$).  A
remarkable fact (emphasized in Section IV of \cite{JS98}) is, however, that the
complete combination of dangerous integrals appearing in $\widetilde{H}_{\rm 
3PN}$
is such that all the pole terms cancel:  $\sum Z_1=0$.  This also implies that
the logarithms depending on the `ultraviolet' regularizing length $l_1$ cancel
in the Hamiltonian \cite{JS98}.  In other words, the combination of dangerous
integrals appearing in $\widetilde{H}_{\rm 3PN}$ is either `finite' (convergent) 
or
globally of a `non dangerous' type.  This is good news, but this still leaves an
ambiguity in the finite value of $\sum I_1(\epsilon_1)$ because it was noticed
in \cite{JS98} that the final regularized result depends on the way one writes
the `bare' integrand $F({\bf x},{\bf x}_1,{\bf x}_2)$ (when transforming it
using standard rules:  integration by parts, and Leibniz's rule).  In addition
to this ambiguity linked to a subset of divergent integrals, there are also
ambiguities in the `contact' terms in the Hamiltonian (of the type $\int
d^3x\,\delta({\bf x}-{\bf x}_1)S({\bf x},{\bf x}_1,{\bf x}_2)$).  One needs to
regularize the (in general singular) limit $\lim_{{\bf x}\to{\bf x}_1}S({\bf
x},{\bf x}_1,{\bf x}_2)$.  In \cite{JS98} one used an Hadamard-like
`partie-finie' prescription (${\rm Pf}_1$) for doing that.  This prescription
is, however, ambiguous at 3PN, notably because ${\rm Pf}_1(\phi_{(2)}^4)\ne[{\rm
Pf}_1(\phi_{(2)})]^4$ \cite{JS99}, where $\phi_{(2)}$ denotes the Newtonian
potential.  As said above, the final ambiguities in the Hamiltonian concern only
two quite specific types of terms, parametrized by $\oms$ and $\omk$ in Eq.\
(\ref{36}).

In Appendix A of this paper, we discuss in more detail these ambiguities, and,
we try to estimate what are the corresponding plausible ranges of values of the
two ambiguity parameters $\oms$ and $\omk$.  For instance, when transforming
(e.g.\ by integration by parts) the divergent integrands we generate some
(rational) values for $\oms$ and $\omk$.  By doing this in various ways, we get
an idea of the range of values that these parameters are likely to be in.  The
final result of Appendix A below is
\be
\label{310}
-10 \lesssim \oms \lesssim 10,\quad
-10 \lesssim \omk \lesssim 10.
\ee
As the symmetric mass ratio $\nu\equiv m_1m_2/(m_1+m_2)^2$ ranges between 0 and 
$1/4$, the ranges (\ref{310}) imply that 
$-3.125\lesssim\sigma(\nu)\lesssim3.125$.  In the following, we shall therefore 
consider as {\it fiducial} range of variation for the (not yet known) 
combination $\sigma$ the simple range
\be
\label{311}
-3 \le -12\nu \le \sigma(\nu) \le 12\nu \le 3.
\ee

\section{Dynamical invariants of the 3PN Hamiltonian}

In this section we derive a complete (and, in fact, overcomplete) set of {\it
invariant} functions of the 3PN dynamics.  Only such functions enter the
principal {\it observables} that can be measured from infinity (say, by
gravitational wave observations).  However, not all the invariants we derive
here can be considered as being directly observable.  In a subsequent paper, we
shall discuss in detail one of the most important observable:  the Last Stable
(circular) Orbit.

We follow closely a work of Damour and Sch\"afer \cite{DS88} in which they
derived the invariants of the 2PN dynamics.  One first uses the invariance of
the Hamiltonian under time translations and spatial rotations.  This yields the
conserved quantities
\be
\label{41}
E \equiv \frac{{\cal E}^{\text{NR}}}{\mu}
\equiv \widehat{H}^{\text{NR}}\left({\bf r},{\bf p}\right),\quad {\bf j} \equiv
\frac{\bf J}{\mu GM} = {\bf r}\times{\bf p}.
\ee
Here ${\cal E}^{\text{NR}}$ denotes the total `non relativistic' energy 
(without rest-mass contribution), and ${\bf J}$ the total angular momentum of 
the binary system in the center of mass frame. Using the Hamilton-Jacobi 
approach, the motion in the plane of the relative trajectory is obtained, in 
polar coordinates $(r,\phi)$, by separating the variables $\widehat{t}\equiv 
t/(GM)$ and $\phi$ in the reduced planar action
\be
\label{42}
\widehat{S} \equiv \frac{S}{\mu GM}
= -E\widehat{t} + j\phi + \int dr\sqrt{R(r,E,j)}.
\ee
Here $j\equiv|{\bf j}|$ and $R(r,E,j)$ is an `effective potential' for the 
radial motion which is obtained by solving the first equation (\ref{41}) for 
$p_r^2\equiv \np^2$ after having replaced $\pp$ by
\be
\label{43}
\pp \equiv \np^2 + ({\bf n}\times{\bf p})^2 = p_r^2 + \frac{j^2}{r^2}.
\ee

Working iteratively in the small parameter $c^{-2}$, and consistently 
neglecting all terms ${\cal O}(c^{-8})$, one finds that $R(r,E,j)$ is given, at 
3PN order, by the following seventh-degree polynomial in $1/r$:
\be
\label{44}
R(r,E,j) = A + \frac{2B}{r} + \frac{C}{r^2} + \frac{D_1^{\text{1PN}}}{r^3}
+ \frac{D_2^{\text{2PN}}}{r^4} + \frac{D_3^{\text{2PN}}}{r^5}
+ \frac{D_4^{\text{3PN}}}{r^6} + \frac{D_5^{\text{3PN}}}{r^7}.
\ee
The coefficients $A$, $B$, $C$ start at Newtonian order, while the extra terms 
$D_i/r^{i+2}$ start at the PN order indicated as superscript. All the 
coefficients $A$, $B$, $C$, $D_i$ are polynomials in $E$ and $j^2$. Their 
explicit expressions are given in Appendix B.

The Hamilton-Jacobi theory shows that the observables of the 3PN motion are 
deducible from the (reduced) `radial action integral'
\be
\label{45}
i_r(E,j) \equiv \frac{2}{2\pi} \int_{r_{\text{min}}}^{r_{\text{max}}} dr 
\sqrt{R(r,E,j)}.
\ee
For instance, the periastron-to-periastron period $P$ and the periastron 
advance $\Delta\Phi\equiv\Phi-2\pi$ per orbit are obtained by differentiating 
the radial action integral:
\bea
\label{46}
\frac{P}{2\pi GM} = \frac{\pa}{\pa E}i_r(E,j),
\\[2ex]
\label{47}
\frac{\Phi}{2\pi} = 1 + \frac{\Delta\Phi}{2\pi} = -\frac{\pa}{\pa j}i_r(E,j).
\eea

To evaluate the invariant function $i_r(E,j)$ we follow Ref.\ \cite{DS88}, 
which was itself following a method devised by Sommerfeld \cite{Sommerfeld} 
within the context of the `old theory of quanta'. This method is explained in 
detail in the Appendix B of Ref.\ \cite{DS88}. The basic idea is the following: 
define $Q(r)\equiv A+2B/r+C/r^2$ and consider the integral
\be
\label{48}
I(\epsilon) \equiv \frac{2}{2\pi} \int_{r_1(\epsilon)}^{r_2(\epsilon)}dr
\sqrt{Q(r) + \epsilon\sum_n \frac{D_n}{r^{n+2}}}.
\ee
Here $\epsilon$ is a formal expansion parameter (actually in the final 
calculations, one takes into account the fact
 that the higher $D_n$'s are multiplied by 
$\epsilon^2$ or $\epsilon^3$, with $\epsilon\sim 1/c^2$ being in factor of 
$D_1^{\text{1PN}}$) and one wishes to compute the expansion of $I(\epsilon)$ as 
$\epsilon\to0$:
$I(\epsilon)
= I_0+\epsilon I_1+\epsilon^2 I_2+\epsilon^3 I_3+{\cal O}(\epsilon^4)$. To 
start with, the limits of integration, $r_1(\epsilon)$ and $r_2(\epsilon)$,
 are the two exact 
($\epsilon$-dependent) real roots of $R(r)=Q(r)+\epsilon\sum D_n/r^{n+2}$. [We 
work in the case where $A<0$, $B>0$, and $C<0$.] The idea is to consider the 
function $\sqrt{R(r)}$ as a complex analytic function defined in a suitably cut 
complex $r$-plane: with, say, a cut along the real segment linking 
$r_1(\epsilon)$ to $r_2(\epsilon)$, and additional cuts from, say, $-\infty$ to 
the other roots. This allows one to rewrite the real integral (\ref{48}) as a 
contour integral
\be
\label{49}
I(\epsilon) = \frac{1}{2\pi} \oint_C dr \sqrt{R(r,\epsilon)},
\ee
where one should note the different numerical prefactor, and where $C$ denotes 
any (counterclockwise, simply winding) contour enclosing the two real roots 
$r_1(\epsilon)$, $r_2(\epsilon)$, and avoiding all the cuts. [The phases of 
$r-r_1$ and $r-r_2$ are defined to be zero when $r$ lies on the real axis on 
the right of $r_2$, while the phase of $\sqrt{A}$ is defined as $+\pi/2$.] If 
we keep the contour $C$ away from the segment $[r_1(\epsilon),r_2(\epsilon)]$, 
we can now directly expand the integrand $\sqrt{R(r,\epsilon)}$ in the contour 
integral (\ref{49}) in powers of $\epsilon$.  This generates the successive 
terms of the expansion $I(\epsilon)
= I_0+\epsilon I_1+\epsilon^2 I_2 +\ldots$, with each term $I_n$ being given
 by a contour integral made of a sum of contributions of the form
\be
\label{410}
J_{q,m} = \frac{1}{2\pi} \oint_C dr
\left(A+\frac{2B}{r}+\frac{C}{r^2}\right)^{\frac{1}{2}-q} \frac{1}{r^m}.
\ee
Each basic integral $J_{q,m}$ appears in the expansion of $I$ multiplied by a 
coefficient which is a polynomial in the $D_n$'s. [At any given PN order, there 
are only a finite number of integrals $J_{q,m}$ to compute; see below.]

There are then two ways to compute the $J_{q,m}$'s. The simplest (the one 
advocated by Sommerfeld) is to expand out the contour $C$ (away from the 
natural cut $[r_1(0),r_2(0)]$ associated to the `unperturbed' quadratic form 
$Ar^2+2Br+C$) until it is deformed into: (i) a {\it clockwise} contour $C_0$ 
around the origin $r=0$, and (ii) {\it anticlockwise} contour $C_{\infty}$ at 
infinity. [Here, one considers that the $J$-integrals are defined in a newly 
cut plane, with a cut only along the segment $[r_1(0),r_2(0)]$.] The value of 
$J_{q,m}$ is then simply given by applying (twice) Cauchy's residue theorem:
it is enough to read off the coefficients of $1/r$ in the two Laurent 
expansions of the integrand of Eq.\ (\ref{410}) for $r\to0$ and $r\to\infty$ 
(keeping track of phases and signs). The other way to compute the $J$'s 
consists in shrinking down the contour $C$ onto the real axis, so as to get 
(twice) a real integral from $r_1(0)+\eta$ to $r_2(0)-\eta$ (with $\eta>0$), 
plus two circular integrals around $r_1(0)$ and $r_2(0)$. As shown in Appendix 
B of \cite{DS88} one can then prove that the $\epsilon$-expansion
(${\rm Exp}_\epsilon$) of $I(\epsilon)$ is simply given by
\be
\label{411}
{\rm Exp}_\epsilon(I(\epsilon))
= {\rm Pf}\left[\frac{2}{2\pi}\int_{r_1(0)}^{r_2(0)} dr\,
{\rm Exp}_\epsilon\Biglb(\sqrt{R(r,\epsilon)}\Bigrb)\right],
\ee
where Pf denotes Hadamard's partie finie of the real $r$-integral. Each 
integral appearing in the formal $\epsilon$-expansion of $\sqrt{R(r,\epsilon)}$ 
on the right-hand side of (\ref{411}) is again the same combination of 
$J$-integrals as above, but now the $J$-integrals are real integrals along the 
segment $[r_1(0),r_2(0)]$, of which one must extract their `partie finie'. The 
parties finies of all those real integrals are easily computed, e.g., by using 
any simple trigonometric parametrization of the radial variable 
($r=a(1-e^2)/(1+e\cos v)$ or $r=a(1-e\cos u)$) to compute the indefinite 
integrals $\int dr [Q(r)]^{\frac{1}{2}-q}r^{-m}$.

The only integrals that one needs to compute at 3PN order are $J_{q,m}$ with 
$0\le q\le3$: when $q=0$, $m=0$; when $q=1$, $3\le m\le7$; when $q=2$, $6\le 
m\le8$; and when $q=3$, $m=9$. We have computed these integrals by the two 
methods indicated above, and checked that the results agree. The results are 
given in Appendix B below.

Finally, one must reexpand all the coefficients $A$, $B$, $C$, $D_n$ (which 
contain various orders in $1/c^2$) to get the complete 3PN expansion of the 
radial action $i_r(E,j)$. The final result reads
\bea
\label{412}
i_r(E,j) &=& -j \bigg\{ 1 - \frac{1}{c^2} \frac{3}{j^2}
- \frac{1}{c^4} \left[ \frac{1}{4}(35-10\nu)\frac{1}{j^4}
+ \frac{1}{2}(15-6\nu)\frac{E}{j^2} \right]
\nonumber\\[2ex]&&
- \frac{1}{c^6} \left[ 
\frac{3}{2}\biglb(i_1(\nu)-\sigma(\nu)\bigrb)\frac{1}{j^6}
+ \biglb(i_2(\nu)-\sigma(\nu)\bigrb)\frac{E}{j^4}
+ 3i_3(\nu)\frac{E^2}{j^2}
\right] \bigg\}
\nonumber\\[2ex]&&
+ \frac{1}{\sqrt{-2E}} \bigg\{ 1 + \frac{1}{c^2} \frac{1}{4}(15-\nu)\,E
+ \frac{1}{c^4} \frac{1}{32}(35+30\nu+3\nu^2)\,E^2
- \frac{1}{c^6} i_4(\nu)\,E^3 \bigg\},
\eea
where
\begin{mathletters}
\label{413}
\bea
\label{i1}
i_1(\nu) &=& \frac{77}{2} + \left(\frac{41}{64}\pi^2-\frac{125}{3}\right)\nu
+ \frac{83}{24}\nu^2,
\\[2ex]
\label{i2}
i_2(\nu) &=& \frac{105}{2} + \left(\frac{41}{64}\pi^2-\frac{218}{3}\right)\nu
+ \frac{221}{24}\nu^2,
\\[2ex]
\label{i3}
i_3(\nu) &=& \frac{1}{4}(5-5\nu+4\nu^2),
\\[2ex]
\label{i4}
i_4(\nu) &=& \frac{1}{128}(21-105\nu+15\nu^2+5\nu^3).
\eea
\end{mathletters}
The differentiation (\ref{46}) then leads to the following expression for the 
periastron-to-periastron period $P$
\bea
\label{414}
\frac{P}{2\pi GM} &=& \frac{1}{(-2E)^{3/2}} \left\{
1 - \frac{1}{c^2} \frac{1}{4}(15-\nu)\,E
\right.\nonumber\\[2ex]&&
+ \frac{1}{c^4} \left[\frac{3}{2}(5-2\nu)\frac{(-2E)^{3/2}}{j}
-\frac{3}{32}(35+30\nu+3\nu^2)\,E^2 \right]
\nonumber\\[2ex]&&\left.
+ \frac{1}{c^6} \left[
\biglb(i_2(\nu)-\sigma(\nu)\bigrb) \frac{(-2E)^{3/2}}{j^3}
- 3i_3(\nu) \frac{(-2E)^{5/2}}{j}
+ 5i_4(\nu)\,E^3 
\right]
\right\},
\eea
where $i_2(\nu)$, $i_3(\nu)$, and $i_4(\nu)$ are given in Eqs.\ (\ref{i2}), 
(\ref{i3}), and (\ref{i4}), respectively. Similarly, the differentiation 
(\ref{47}) yields, for the dimensionless parameter
\be
\label{415}
k \equiv \frac{\Delta\Phi}{2\pi} = \frac{\Phi-2\pi}{2\pi}
\ee
measuring the fractional periastron advance per orbit, the result
\be
\label{416}
k = \frac{1}{c^2} \frac{3}{j^2} \left\{
1 + \frac{1}{c^2} \left[ \frac{5}{4}(7-2\nu)\frac{1}{j^2}
+ \frac{1}{2}(5-2\nu)\,E \right]
+ \frac{1}{c^4} \left[
\frac{5}{2}\biglb(i_1(\nu)-\sigma(\nu)\bigrb)\frac{1}{j^4}
+ \biglb(i_2(\nu)-\sigma(\nu)\bigrb)\frac{E}{j^2} + 3i_3(\nu)\,E^2
\right] \right\},
\ee
where $i_1(\nu)$, $i_2(\nu)$, and $i_3(\nu)$ are given in Eqs.\ (\ref{i1}), 
(\ref{i2}), and (\ref{i3}), respectively.

We shall also follow Ref.\ \cite{DS88} in giving the Hamiltonian as a function 
of the Delaunay (reduced) action variables
\be
\label{417}
n \equiv i_r + j = \frac{\cal N}{\mu GM},\quad
j = \frac{J}{\mu GM},\quad
m \equiv j_z = \frac{J_z}{\mu GM}.
\ee
In the quantum language, ${\cal N}/\hbar$ is the principal quantum number, 
$J/\hbar$ the total angular-momentum quantum number, and $J_z/\hbar$ the 
magnetic quantum number. [They are adiabatic invariants of the dynamics and they 
are (approximately) quantized in integers.] By rotational symmetry, the 
(reduced) magnetic quantum number $m$ does not enter the Hamiltonian. By 
inverting Eq.\ (\ref{412}) one finds
\bea
\label{418}
\widehat{H}(n,j,m) &=& - \frac{1}{2n^2} \Bigg\{ 1 +
\frac{1}{c^2} \bigg[ \frac{6}{j n}-\frac{1}{4}(15-\nu)\frac{1}{n^2} \bigg]
\nonumber\\[2ex]&&
+ \frac{1}{c^4} \bigg[ \frac{5}{2}(7-2\nu)\frac{1}{j^3 n}
+ \frac{27}{j^2 n^2}
- \frac{3}{2}(35-4\nu)\frac{1}{j n^3}
+ \frac{1}{8}(145-15\nu+\nu^2)\frac{1}{n^4} \bigg]
\nonumber\\[2ex]&&
+ \frac{1}{c^6} \bigg[
3\biglb(i_1(\nu)-\sigma(\nu)\bigrb)\frac{1}{j^5 n}
+ \frac{45}{2}(7-2\nu)\frac{1}{j^4 n^2}
- \biglb(i_5(\nu)-\sigma(\nu)\bigrb)\frac{1}{j^3 n^3}
- \frac{45}{2}(20-3\nu)\frac{1}{j^2 n^4}
\nonumber\\[2ex]&&
+ \frac{3}{2}(275-50\nu+4\nu^2)\frac{1}{j n^5}
- \frac{1}{64}(6363-805\nu+90\nu^2-5\nu^3)\frac{1}{n^6}
\bigg] \Bigg\},
\eea
where
\be
\label{i5}
i_5(\nu) = \frac{303}{4} + \left(\frac{41}{64}\pi^2-\frac{1427}{12}\right)\nu
+ \frac{281}{24}\nu^2.
\ee
The results (\ref{414}), (\ref{416}), and (\ref{418}) confirm the 2PN results of 
\cite{DS88} and extend them to the 3PN level.

The angular frequencies of the (generic) rosette motion of the binary system are 
then given by differentiating $\widehat{H}$ with respect to the action 
variables. 
Namely,
\bea
\label{419}
\omega_{\text{radial}} = \frac{2\pi}{P}
= \frac{1}{GM} \frac{\pa\widehat{H}(n,j,m)}{\pa n},
\\[2ex]
\label{420}
\omega_{\text{periastron}} = \frac{\Delta\Phi}{P} = \frac{2\pi k}{P}
= \frac{1}{GM} \frac{\pa\widehat{H}(n,j,m)}{\pa j}.
\eea
Here, $\omega_{\text{radial}}$ is the angular frequency of the radial motion, 
i.e., the angular frequency of the return to the periastron, while 
$\omega_{\text{periastron}}$ is the average angular frequency with which the 
major axis advances in space.

It is interesting to note that the ambiguity parameter $\sigma(\nu)$ enters only 
in two of the six independent combinations of $n$ and $j$ which enter at the 3PN 
level, cf.\ Eq.\ (\ref{418}). We note also that if we consider the fiducial 
range (\ref{311}) the numerical effect of $\sigma(\nu)$ seems to remain rather 
small compared to the `non ambiguous' contribution.
Indeed, $\sigma(\nu)$ enters Eqs.\ (\ref{412}), (\ref{414}), (\ref{416}), and 
(\ref{418}) only through the differences $i_1(\nu)-\sigma(\nu)$, 
$i_2(\nu)-\sigma(\nu)$, and $i_5(\nu)-\sigma(\nu)$; $i_1(\nu)$ varies between 
38.50 when $\nu=0$, and 29.88 when $\nu=1/4$, $i_2(\nu)$ varies between 52.50 
when $\nu=0$, and 36.49 when $\nu=1/4$, while $i_5(\nu)$ varies between 75.75 
when $\nu=0$, and 48.33 when $\nu=1/4$. The $\sigma$ ambiguity therefore has 
only a limited fractional effect on the coefficients it influences: $\le10.0\%$, 
$\le8.2\%$, and $\le6.2\%$ for the coefficients influenced through 
$i_1(\nu)-\sigma(\nu)$, $i_2(\nu)-\sigma(\nu)$, and $i_5(\nu)-\sigma(\nu)$, 
respectively.

By contrast we wish to emphasize that the use of the approximate ansatz (\`a la 
Wilson-Mathews \cite{WM95}) reducing the spatial metric to being conformally 
flat affects in a numerically much more drastic way the dynamical invariants.
 Indeed, the 
difference starts at the observationally much more significant 2PN order. The 
Wilson-Mathews truncation of the 2PN Hamiltonian is simply obtained by setting 
${\htt ij}=0$ in the formulas of \cite{S85}. It reads (in ADM coordinates)
\bea
\label{421}
\widehat{H}_{\rm WM}({\bf r},{\bf p})
&=& \frac{\pp}{2} - \frac{1}{r}
+ \frac{1}{c^2} \left\{
\frac{1}{8}(3\nu-1)\ppp^2
- \frac{1}{2}\left[(3+\nu)\pp+\nu\np^2\right]\frac{1}{r}
+ \frac{1}{2r^2} \right\}
\nonumber\\[2ex]&&
+ \frac{1}{c^4} \left\{
\frac{1}{16}\left(1-5\nu+5\nu^2\right)\ppp^3
+ \frac{5}{8}\left(1-4\nu\right)\ppp^2\frac{1}{r}
\right.\nonumber\\[2ex]&&\left.
+ \frac{1}{4} \left[(10+19\nu)\pp-3\nu\np^2\right]\frac{1}{r^2}
- \frac{1}{4}(1+\nu)\frac{1}{r^3}
\right\}.
\eea
Rieth \cite{R97} has computed (following, as above, the route of Ref.\ 
\cite{DS88}) the invariant functions of $\widehat{H}_{\rm WM}$, Eq.\ 
(\ref{421}). 
They read (see Eqs.\ (5) and (8) in \cite{R97})
\bea
\label{422}
\frac{P}{2\pi GM} &=& \frac{1}{(-2E)^{3/2}} \left\{
1 - \frac{1}{c^2} \frac{1}{4}(15-\nu)\,E
+ \frac{1}{c^4} \left[ \frac{1}{8}(60-18\nu-41\nu^2)\frac{(-2E)^{3/2}}{j}
-\frac{15}{32}(7+6\nu-25\nu^2)\,E^2 \right] \right\},
\\[2ex]
\label{423}
k &=& \frac{1}{c^2} \frac{3}{j^2} \left\{
1 + \frac{1}{c^2} \left[ \frac{1}{16}(140-54\nu-31\nu^2)\frac{1}{j^2}
+ \frac{1}{24}(60-18\nu-41\nu^2)\,E \right] \right\}.
\eea
The corresponding Delaunay Hamiltonian reads
\bea
\label{424}
\widehat{H}_{\rm WM}(n,j,m) &=& - \frac{1}{2n^2} \Bigg\{ 1 +
\frac{1}{c^2} \bigg[ \frac{6}{j n}-\frac{1}{4}(15-\nu)\frac{1}{n^2} \bigg]
\nonumber\\[2ex]&&
+ \frac{1}{c^4} \bigg[ \frac{1}{8}(140-54\nu-31\nu^2)\frac{1}{j^3 n}
+ \frac{27}{j^2 n^2}
- \frac{1}{8}(420-42\nu-41\nu^2)\frac{1}{j n^3}
+ \frac{5}{8}(29-3\nu-3\nu^2)\frac{1}{n^4} \bigg] \Bigg\}.
\eea

When comparing Eqs.\ (\ref{422})--(\ref{424}) to the 2PN-accurate versions of
Eqs.\ (\ref{414}), (\ref{416}), and (\ref{418}) we see that the maximum
fractional difference in the 2PN-level coefficients (between the exact
$\widehat{H}_{\rm 2PN}$ and the Wilson-Mathews truncation $\widehat{H}_{\rm
WM}$) is about 23\% (for $\nu=1/4$, and the coefficient of $ E^2 $ on the R.H.S.
of Eq.\ (\ref{422})).  2PN-level effects are crucial in determining important
observables, such as the location of the Last Stable Orbit (as discussed, e.g.,
in Refs.\ \cite{DIS98,BD99}, and in our follow up work \cite{DJS2}), and are
numerically more important than 3PN-level effects (even near the Last Stable
Orbit the PN expansion parameter $v^2/c^2 \sim 1/6$ is still smallish).
Therefore we consider that the Wilson-Mathews truncation is, {\it a priori},
unacceptably inaccurate.  We shall come back below, and in Ref.\ \cite{DJS2}, to
the more delicate problem of judging whether the 3PN-level uncertainty
(\ref{311}) on $\sigma$ is practically acceptable or not.

\section{Dynamical invariants for the 3PN circular motion}

In the previous section we considered the invariant functions associated to 
generic (bound) 3PN orbits ($E<0$). These orbits correspond to a doubly periodic 
rosette motion. The case of circular orbits, though quite special from a general 
point of view, plays a physically very important role and deserves a dedicated 
study. Circular orbits represent the degenerate limit where the integration 
range $[r_{\text{min}},r_{\text{max}}]$ in Eq.\ (\ref{45}) shrinks to a point. 
[In invariant terms, circular orbits are characterized by the fact that the 
motion becomes simply periodic, instead of doubly periodic.] Technically they 
are therfore characterized by setting the radial action $i_r$, Eq.\ (\ref{45}), 
to zero. In other words, the reduced `principal quantum number' $n=i_r+j$ should 
be equalled to $j$. Using Eq.\ (\ref{418}) this gives the following link, 
between 
energy and angular momentum, characterizing circular orbits:
\be
\label{51}
E_{\text{circ}} = -\frac{1}{2j^2} \Bigg[ 1
+ \frac{1}{c^2} \frac{1}{4}(9+\nu)\frac{1}{j^2}
+ \frac{1}{c^4} \frac{1}{8}(81-7\nu+\nu^2)\frac{1}{j^4}
+ \frac{1}{c^6} 2\biglb(o_1(\nu)-\sigma(\nu)\bigrb)\frac{1}{j^6} \Bigg],
\ee
where
\be
\label{51a}
o_1(\nu) = \frac{3861}{128} + \frac{1}{64}\left(41\pi^2-\frac{8833}{6}\right)\nu
+ \frac{313}{192}\nu^2 + \frac{5}{128}\nu^3.
\ee
Note that, evidently, it was not needed to derive the full Delaunay Hamiltonian 
(\ref{418}) to get the result (\ref{51}). It would have been sufficient to 
impose that the effective radial potential $R(r,E,j)$, Eq.\ (\ref{44}), have a 
double zero, i.e.\ that ${\pa R}/{\pa r}=0={\pa^2 R}/{\pa r^2}$, or 
equivalently that the Hamiltonian $\widehat{H}({\bf r},{\bf p})$ (reduced to 
circular orbits by setting ${\bf n}\cdot{\bf p}=0$, $\pp=j^2/r^2$) has a minimum 
when considered as a function of $r$ for fixed $j$. Let us also give (though it 
is not an invariant function) the expression of the coordinate radius $r$ (which 
is the shifted $r'$ of Eq.\ (\ref{223})) in function of $j$
\be
r = j^2 \bigg\{ 1 - \frac{1}{c^2} \frac{4}{j^2}
- \frac{1}{c^4} \frac{1}{8}(74-43\nu)\frac{1}{j^4}
- \frac{1}{c^6} \left[ 55
+ \left(\frac{163}{64}\pi^2-\frac{1655}{16}-4\oms\right)\nu
+ \left(\frac{61}{6}-5\omk\right)\nu^2 \right] \frac{1}{j^6} \bigg\}.
\ee

An important observational quantity is the angular frequency of circular orbits, 
$\omega_{\text{circ}}$. Let us first recall that $\omega_{\text{circ}}$ differs 
from the above considered periastron-to-periastron or radial frequency 
$\omega_{\text{radial}}$, Eq.\ (\ref{419}). In fact, because the frequency 
$\omega_{\text{radial}}$ measures the frequency in the `rotating frame' of an 
orbit which precesses in space with the frequency $\omega_{\text{periastron}}$ 
(\ref{420}), it is easy to see that
\be
\label{52}
\omega_{\text{circ}} = \omega_{\text{radial}} + \omega_{\text{periastron}}
= 2\pi \frac{1+k}{P}.
\ee
Remembering that $n=j$ along circular orbits we see from Eq.\ (\ref{419}) that
\be
\label{53}
\omega_{\text{circ}} = \frac{1}{GM} \frac{d E_{\text{circ}}}{dj},
\ee
which corresponds to the usual link $d{\cal E}=\omega_{\text{circ}}dJ$ between 
the (unreduced) total energy ${\cal E}=M c^2+{\cal E}^{\text{NR}}$ and the total 
angular momentum $J=\mu GMj$. As usual, it is convenient to introduce the 
(coordinate-invariant) dimensionless variable
\be
\label{54}
x \equiv \left(\frac{GM\omega_{\text{circ}}}{c^3}\right)^{2/3}.
\ee
In the following, we shall set $c=1$ to simplify the formulas.
Inserting Eq.\ (\ref{51}) into Eq.\ (\ref{53}) gives the link
\be
\label{55}
GM\omega_{\text{circ}} \equiv x^{3/2} = \frac{1}{j^3} \Bigg[
1 + \frac{1}{2}(9+\nu)\frac{1}{j^2}
+ \frac{3}{8}(81-7\nu+\nu^2)\frac{1}{j^4}
+ 8\biglb(o_1(\nu)-\sigma(\nu)\bigrb)\frac{1}{j^6} \Bigg].
\ee
By inverting this relation one gets $j$ as a function of $x$:
\be
\label{56}
j_{\text{circ}} = x^{-1/2} \left[ 1 + \frac{1}{6}(9+\nu) x
+ \frac{1}{24}(81-57\nu+\nu^2) x^2
+ \frac{8}{3}\biglb(o_2(\nu)-\sigma(\nu)\bigrb) x^3
\right],
\ee
where
\be
\label{56a}
o_2(\nu) = \frac{405}{128} + \frac{1}{64}\left(41\pi^2-\frac{6889}{6}\right)\nu
+ \frac{421}{192}\nu^2 + \frac{7}{3456}\nu^3.
\ee

At this point it is useful to consider the test-mass limit ($\nu\to0$). Indeed, 
in this limit, i.e.\ in the case of a test particle $m_2$ moving, on a 
circular orbit, in a 
Schwarzschild background of mass $m_1$ (with $m_2\ll m_1$), we have the 
relations 
(see, e.g., Eqs.\ (5.19)--(5.22) in \cite{BD99})
\begin{mathletters}
\bea
\label{57a}
j &=& \frac{1}{\sqrt{x(1-3x)}},
\\[2ex]
\label{57b}
\frac{1}{j^2} &=& x(1-3x).
\eea
\end{mathletters}
Here the variable $x$ is defined as in Eq.\ (\ref{54}), with $M=m_1+m_2$, and 
$j\equiv J/(\mu GM)=J/(Gm_1m_2)$, where $J$ is the total angular momentum of the 
system (which is entirely carried by the test particle). [At this stage we could 
equivalently consider that $M$ is the mass of the black hole and $\mu\equiv 
m_1m_2/M$ the mass of the test particle.] The test mass result (\ref{57b}) 
suggests to focus on the function $j^{-2}(x)$, which, from Eq.\ (\ref{56}), is 
given, when $\nu\ne0$, by
\be
\label{58}
\frac{1}{j_{\text{circ}}^2} = x \left[ 1 - \frac{1}{3}(9+\nu) x
+ \frac{25}{4}\nu x^2
- \frac{16}{3}\biglb(o_3(\nu)-\sigma(\nu)\bigrb) x^3
\right],
\ee
where
\be
\label{58a}
o_3(\nu) = \frac{1}{64}\left(41\pi^2-\frac{5269}{6}\right)\nu
+ \frac{511}{192}\nu^2 - \frac{1}{432}\nu^3.
\ee
Let us also consider the energy for circular orbits. By inserting Eq.\ 
(\ref{58}) into Eq.\ (\ref{51}) we get
\be
\label{59}
E_{\text{circ}} \equiv \frac{{\cal E}^{\text{total}}-M}{\mu}
= -\frac{1}{2}x \left[1 - \frac{1}{12}(9+\nu)x
- \frac{1}{24}(81-57\nu+\nu^2)x^2
- \frac{10}{3}\biglb(o_2(\nu)-\sigma(\nu)\bigrb)x^3
\right],
\ee
where $o_2(\nu)$ is given in Eq.\ (\ref{56a}).
It is easily verified that Eqs.\ (\ref{56}) and (\ref{59}) satisfy (at the 3PN 
accuracy) the exact identity following from Eq.\ (\ref{53})
\be
\label{510}
\frac{d E_{\text{circ}}}{dx} = x^{3/2} \frac{dj_{\text{circ}}}{dx}.
\ee

When comparing Eq.\ (\ref{59}) with the test-mass limit, there is a problem of 
definition of the best energy variable to use in a binary system. Indeed, if we 
consider a test mass of mass $m_2$ moving around a Schwarzschild black hole of 
mass $m_1$, the total energy of the system reads
\be
\label{511}
{\cal E}^{\text{total}} = m_1 + {\cal E}^{\text{TM}}_2
= m_1 - k_\mu p_2^\mu
\simeq m_1 - \frac{p_{1\mu}p_2^\mu}{m_1},
\ee
where $p_2^\mu$ is the 4-momentum of the test mass $m_2$, and
${\cal E}^{\text{TM}}_2=-k_\mu p_2^\mu$ is the conserved relativistic energy. 
Here, $k^\mu$ is the time-translation Killing vector which is (approximately) 
$k^\mu=p_1^\mu/m_1$, where $p_1^\mu$ is the 4-momentum of the large mass.
Eq.\ (\ref{511}), and the known results for circular motion in Schwarzschild 
(still with $\mu\equiv m_1m_2/(m_1+m_2)\simeq m_2$ modulo ${\cal O}(\nu)$),
yield
\be
\label{512}
E = \frac{{\cal E}^{\text{total}}-m_1-m_2}{\mu}
\simeq \frac{{\cal E}^{\text{TM}}_2-m_2}{m_2}
= \frac{1-2x}{\sqrt{1-3x}} - 1.
\ee
As emphasized in Ref.\ \cite{DIS98} the expression (\ref{511}) is very 
asymmetric with respect to the labels 1 and 2. It seems much better to consider 
the Mandelstam invariant
$s=({\cal E}^{\text{total}})^2=-(p_1+p_2)^2=m_1^2+m_2^2-2p_1\cdot p_2$ and 
therefore the combination
\be
\label{513}
\frac{({\cal E}^{\text{total}})^2-m_1^2-m_2^2}{2m_1m_2}
= -\frac{p_1\cdot p_2}{m_1m_2}
\simeq \frac{{\cal E}^{\text{TM}}_2}{m_2}
= \frac{1-2x}{\sqrt{1-3x}}.
\ee
This consideration, and the fact that the energy combination defined by the 
left-hand side of (\ref{513}) still exhibits a branch cut singularity in the 
complex $x$-plane, motivated Ref.\ \cite{DIS98} to introduce a new invariant 
energy function $e$, defined by
\be
\label{514}
1 + e \equiv
\left(\frac{({\cal E}^{\text{total}})^2-m_1^2-m_2^2}{2m_1m_2}\right)^2.
\ee
In terms of $E\equiv\widehat{\cal E}^{\text{NR}}
\equiv{\cal E}^{\text{NR}}/\mu\equiv({\cal E}^{\text{total}}-M)/\mu$, the new 
energy function $e$ reads
\be
\label{515}
e = \big(1+E+\frac{\nu}{2}E^2\big)^2 - 1
= 2E + (1+\nu)E^2 + \nu E^3 + \frac{\nu^2}{4}E^4.
\ee
Inserting Eq.\ (\ref{59}) into Eq.\ (\ref{515}) yields (at 3PN)
\be
\label{516}
e(x,\nu) = -x \left[ 1 - \frac{1}{3}(3+\nu)x - \frac{1}{12}(36-35\nu)x^2
- \frac{10}{3}\biglb(o_4(\nu)-\sigma(\nu)\bigrb)x^3 \right],
\ee
where
\be
\label{516a}
o_4(\nu) = \frac{27}{10}
+ \frac{1}{16}\left(\frac{41}{4}\pi^2-\frac{4309}{15}\right)\nu
+ \frac{77}{30}\nu^2 - \frac{1}{270}\nu^3.
\ee
{}From Eq.\ (\ref{513}) the test-mass limit of this function takes the simple 
(meromorphic) form
\be
\label{517}
e(x,0) = -x \frac{1-4x}{1-3x}.
\ee

For completeness, let us note that the test-mass limit of the invariant link 
between energy and angular momentum, Eq.\ (\ref{51}), reads
\be
\label{518}
1 + E = \frac{\sqrt{2}}{3} 
\frac{2+\sqrt{\Delta(j)}}{\sqrt{1+\sqrt{\Delta(j)}}};\quad
\Delta(j) \equiv 1 - \frac{12}{j^2}.
\ee
Actually, the result (\ref{518}) holds only along the sequence of {\it stable} 
circular orbits (of Schwarzschild radius $R>6GM$, i.e.\ for a frequency 
parameter $x=GM/R<1/6$). When $x>1/6$ (but $x<1/3$), i.e.\ for unstable circular 
orbits $\sqrt{\Delta}$ should be replaced by $-\sqrt{\Delta}$ in (\ref{518}). 
This corresponds to the fact that the curve $E=E(j)$ has a cusp at the Last 
Stable Orbit.

To complete our list of exact results in the test-mass limit, let us also note 
that the periastron parameter $1+k=\Phi/(2\pi)$ is given, in the limit of 
circular orbits, by \cite{DS88,JS91}
\be
\label{519}
1 + k = \left(1-\frac{12}{j^2}\right)^{-1/4}.
\ee
Inserting Eq.\ (\ref{51}) into Eq.\ (\ref{423}) we get, for circular orbits,
\be
\label{520}
k_{\text{circ}} = \frac{3}{j^2}
+ \frac{1}{2}(45-12\nu) \frac{1}{j^4}
+ 6\biglb(o_5(\nu)-\sigma(\nu)\bigrb) \frac{1}{j^6},
\ee
where
\be
\label{520a}
o_5(\nu) = \frac{135}{4} + \left(\frac{41}{64}\pi^2-\frac{101}{3}\right)\nu
+ \frac{53}{24}\nu^2.
\ee
{}From Eq.\ (\ref{520}) follows that
\be
\label{521}
(1+k_{\text{circ}})^{-4} = 1 - \frac{12}{j^2} + 24\nu \frac{1}{j^4}
- 24\biglb(o_6(\nu)-\sigma(\nu)\bigrb) \frac{1}{j^6},
\ee
where
\be
\label{521a}
o_6(\nu) = \left(\frac{41}{64}\pi^2-\frac{56}{3}\right)\nu + \frac{53}{24}\nu^2.
\ee

We will use in a further work \cite{DJS2} the various invariant functions
computed above, at the 3PN level, to discuss the observable quantities
associated to the Last Stable (circular) Orbit of a binary system.  For the time
being, we shall only remark that, contrary to what happened above in the
dynamical invariants for generic orbits, one anticipates that the incomplete
knowledge of the ambiguity parameter $\sigma(\nu)$ might be much more serious in
the dynamics of circular orbits.  Indeed, if we consider for instance the
function $j^{-2}(x)$, Eq.\ (\ref{58}), we see that $\sigma(\nu)$ modifies the
coefficient of $x^3$ through the difference $o_3(\nu)-\sigma(\nu)$; $o_3(\nu)$
vanishes when $\nu=0$ and decreases monotonically with $\nu$, taking the value
$-1.683$ when $\nu=1/4$.  Therefore the addition of $\sigma(\nu)$, within the
range (\ref{311}), can modify a lot the coefficient of $x^3$ in $j^{-2}(x)$.
Similarly, in the case of the original energy function $E(x)$, Eq.\ (\ref{59}),
$\sigma(\nu)$ is subtracted from the coefficient $o_2(\nu)$ which decreases
monotonically with $\nu$ and varies between $o_2(0)=3.164$ and $o_2(1/4)=0.397$. 
This issue will be discussed in detail in \cite{DJS2}.

\section*{Acknowledgments}

This work was supported in part by the KBN Grant No.\ 2 P03B 094 17 (to P.J.)  
and Max-Planck-Gesellschaft Grant No.\ 02160-361-TG74 (to G.S.).
P.J.\ and G.S.\ thank the Institut des Hautes \'Etudes Scientifiques for 
hospitality during crucial stages of the collaboration.

\appendix

\section{Ambiguities in the 3PN ADM two-body point-mass Hamiltonian}

In this appendix we recalculate the 3PN ADM two-body point-mass Hamiltonian
given in Ref.\ \cite{JS98} and we discuss the plausible range of the ambiguity
parameter $\sigma$, Eq.\ (\ref{37}).  [We use here the notation of \cite{JS98},
notably $ 16\pi G=1$.]  The motivation for recalculating the Hamiltonian is
three-fold:  (i) We want to avoid using explicit solutions of some Poisson
equations with complicated (and singular) source terms (it was shown in Ref.\
\cite{JS99} that it may not be possible to obtain unique solutions to these
equations).  (ii) Before applying any regularization procedure we want to
represent the Hamiltonian as a volume integral with an integrand which diverges
locally as weakly as possible (thus absorbing stronger divergent terms into
surface integrals).  (iii) We wish to eliminate all `contact' terms (Dirac
delta functions) in the Hamiltonian, so that {\em only one} regularization
procedure is needed (the one used to regularize the volume integral giving the
Hamiltonian).

Our starting point is the 3PN Hamiltonian $\widetilde{H}_{\rm 3PN}$ as given by 
Eqs.\ (54)--(60) of Ref.\ \cite{JS98}. In the first step we eliminate from 
$\widetilde{H}_{31}$ (given by Eq.\ (55) of \cite{JS98}) the function 
$\phi_{(8)1}$ by integrating by parts\footnote{In this appendix we abbreviate 
$\delta({\bf x}-{\bf x}_a)$ by $\delta_a$.}:
$\phi_{(8)1}\sum_a m_a\delta_a=-\phi_{(8)1}\left(\Delta\phi_{(2)}\right)
=-\left(\Delta\phi_{(8)1}\right)\phi_{(2)}+$ exact divergence. To do this we use 
the equation (see Eqs.\ (A11) and (A12) in \cite{JS98})
\bea
\label{a1}
\Delta\phi_{(8)1} &=& \sum_a\left\{
\frac{1}{512}\phi_{(2)}^3-\frac{1}{32}\phi_{(2)}\phi_{(4)}
+\frac{1}{8}\phi_{(6)}
+\left(\frac{5}{16}\phi_{(4)}-\frac{15}{128}\phi_{(2)}^2\right)
\frac{{\bf p}_a^2}{m_a^2}
\right.\nonumber\\[2ex]&&\left.
-\frac{9}{64}\phi_{(2)}\frac{({\bf p}_a^2)^2}{m_a^4}
-\frac{1}{16}\frac{({\bf p}_a^2)^3}{m_a^6}
+\frac{1}{2}\frac{p_{ai}p_{aj}}{m_a^2}{\httiv ij}\right\}m_a\delta_a
+\frac{1}{2}\phi_{(4),ij}{\httiv ij}
+\frac{1}{2}\Delta\left\{\left({\httiv ij}\right)^2\right\}.
\eea
When performing an integration by parts we always check that the 
exact-divergence terms fall off at infinity faster than $1/r^4$, so that
(formally) they do not 
contribute to the Hamiltonian. In the next step we split the function 
$\phi_{(6)}$ into three parts (see Eq.\ (A4) in \cite{JS98})
\be
\label{a2}
\phi_{(6)}=\phi_{(6)1}+\phi_{(6)2}+\phi_{(6)3},
\ee
and we eliminate the functions $\phi_{(6)2}$ and $\phi_{(6)3}$. To perform this 
we use again integration by parts and equations (see Eqs.\ (A6) and (A7) in 
\cite{JS98})
\bea
\label{a3}
\Delta\phi_{(6)2} &=& -\left({\pitiii ij}\right)^2,
\\[2ex]
\label{a4}
\Delta\phi_{(6)3} &=& \frac{1}{2}\phi_{(2),ij}{\httiv ij}.
\eea
The elimination of the functions $\phi_{(8)1}$, $\phi_{(6)2}$, and $\phi_{(6)3}$ 
means that the goal indicated in the item (i) at the beginning of this appendix 
is achieved; the Hamiltonian depends only on the three lowest-order solutions of 
the Hamiltonian constraint: $\phi_{(2)}$, $\phi_{(4)}$, and 
$\phi_{(6)1}$. These three functions are not influenced by the possible
 ambiguities 
in solving the Hamiltonian constraint discussed in Ref.\ \cite{JS99}.
To achieve the goals of items (ii) and (iii) one must perform many 
integration by parts and employ several times the equations fulfilled by the 
functions $\phi_{(2)}$, $\phi_{(4)}$, and $\phi_{(6)1}$, which read (see Eqs.\ 
(15), (16), and (A5) in \cite{JS98})
\bea
\label{a5}
\Delta\phi_{(2)} &=& -\sum_a m_a\delta_a,
\\[2ex]
\label{a6}
\Delta\phi_{(4)} &=& \sum_a
\left(\frac{1}{8}\phi_{(2)}-\frac{1}{2}\frac{{\bf p}_a^2}{m_a^2}\right) 
m_a\delta_a,
\\[2ex]
\label{a7}
\Delta\phi_{(6)1} &=& \sum_a\left(
-\frac{1}{64}\phi_{(2)}^2+\frac{1}{8}\phi_{(4)}
+\frac{5}{16}\phi_{(2)}\frac{{\bf p}_a^2}{m_a^2}
+\frac{1}{8}\frac{({\bf p}_a^2)^2}{m_a^4} \right)m_a\delta_a.
\eea

Finally, we are able to put the 3PN two-body point-mass ADM Hamiltonian in the 
following form (dropping many surface terms, after checking that they fall off
fast enough at infinity not to contribute to the Hamiltonian)
\be
\label{a8}
\widetilde{H}_{\rm 3PN} = -\frac{5}{128}\sum_a\frac{({\bf p}_a^2)^4}{m_a^7}
+ \int d^3\!x\,(h_1 + h_2 + h_3),
\ee
where
\begin{mathletters}
\label{a9}
\bea
\label{h1}
h_1 &=& \frac{1}{8} 
\left(\frac{5}{16}\phi_{(2)}^3-\frac{11}{4}\phi_{(2)}\phi_{(4)}
+\phi_{(6)1}\right)_{\!\!,i}S_{(4),i} + \frac{1}{16} 
\left(\frac{13}{8}\phi_{(2)}^2-3\phi_{(4)}\right)_{\!\!,i}S_{(4),i}
+ \frac{7}{64}\phi_{(2),i}S_{(8),i}
- \frac{1}{4}{\ghttiv ijk}S_{(6)ij,k}
\nonumber\\[2ex]&&
+ \left(\frac{1}{8}S_{(4)}-\frac{5}{4}\phi_{(4)}\right)
\left({\pitiii ij}\right)^2
+ \left(\frac{1}{16}S_{(4),j}-\frac{3}{8}\phi_{(4),j}\right)
\phi_{(2),i}{\httiv ij}
-\left(\left(\phi_{(2)}{\pitiii ij}\right)^{\rm TT}\right)^2
\nonumber\\[2ex]&&
- \frac{1}{4}\left({\httivdot ij}\right)^2
-\frac{1}{2}{\httivdot ij}\left(\phi_{(2)}{\pitiii ij}\right)^{\rm TT},
\\[2ex]
\label{h2}
h_2 &=& \frac{1}{32} \left[
\left(\phi_{(2)}^2\right)_{\!\!,i}\phi_{(6)1,i}
+ 3\left(\phi_{(2)}\phi_{(4)}\right)_{\!\!,i}\phi_{(4),i}
+ 2\left(\phi_{(4)}^2\right)_{\!\!,i}\phi_{(2),i} \right],
\\[2ex]
\label{h3}
h_3 &=& \frac{35}{64}\phi_{(2)}^2\left({\pitiii ij}\right)^2
+2{\pitiii ik}{\pitiii jk}{\httiv ij}
-2\left(2\pi_{(3)}^i+\Delta^{-1}\pi_{(3),il}^l\right)_{,k}
{\pitiii jk}{\httiv ij}
\nonumber\\[2ex]&&
+\left(2\pi_{(3)}^i+\Delta^{-1}\pi_{(3),il}^l\right)
\left(2\pi_{(3)}^k+\Delta^{-1}\pi_{(3),km}^m\right)_{,j}{\ghttiv ijk}
\nonumber\\[2ex]&&
+\frac{5}{64}\phi_{(2)}\phi_{(2),i}\phi_{(2),j}{\httiv ij}
- \left(\frac{1}{4}{\httiv ij}+\frac{5}{8}S_{(4)ij}\right)_{\!\!,k}
\left(\phi_{(2)}{\httiv ij}\right)_{\!\!,k}.
\eea
\end{mathletters}
In Eqs.\ (\ref{h1})--(\ref{h3}) the following notation was introduced
\begin{mathletters}
\label{a10}
\bea
S_{(4)} &\equiv& -\frac{1}{4\pi} \sum_a\frac{{\bf p}_a^2}{m_a r_a},
\\[2ex]
S_{(4)ij} &\equiv& -\frac{1}{4\pi} \sum_a\frac{p_{ai}p_{aj}}{m_a r_a},
\\[2ex]
S_{(6)} &\equiv& -\frac{1}{4\pi} \sum_a\frac{({\bf p}_a^2)^2}{m_a^3 r_a},
\\[2ex]
S_{(6)ij} &\equiv& -\frac{1}{4\pi}
\sum_a\frac{{\bf p}_a^2p_{ai}p_{aj}}{m_a^3 r_a},
\\[2ex]
S_{(8)} &\equiv& -\frac{1}{4\pi} \sum_a\frac{({\bf p}_a^2)^3}{m_a^5 r_a}.
\eea
\end{mathletters}
We have not found any ambiguity connected with the regularization of the
divergent integrals contained in the first part $h_1$ of the Hamiltonian (see 
below for details). Regularization of the $h_2$ part contributes only to the 
static ambiguity, and regularization of the $h_3$ part contributes only to the 
kinetic ambiguity. The static part of the 3PN Hamiltonian (discussed in detail 
in Ref.\ \cite{JS99}; see footnote \ref{ft:static} in the present paper) is 
equal to $\int{d^3\!x\,h_2}$. 

We regularize all the divergent integrals involved in the Hamiltonian given 
by Eqs.\ (\ref{a8})--(\ref{a9}) by means of the Riesz-formula-based 
regularization technique described in Appendix B 2 of \cite{JS98} and 
rediscussed in Sec.\ III of the present paper. The final result, after 
eliminating the time derivatives of the particle positions and momenta by 
using the Newtonian equations of motion, Eq.\ (\ref{31}), is given in Eq.\ 
(\ref{36}).
 
To investigate the lack of uniqueness of the regularization method we have used 
the following procedure. To many individual terms in Eqs.\ (\ref{a9}) one can 
associate exact divergences which must be added to these terms to replace them 
by some of their integration-by-parts equivalents. E.g., with the first term in 
Eq.\ (\ref{h1}) one can associate two exact divergences, 
$-\frac{5}{128}[(\phi_{(2)}^3)_{,i}S_{(4)}]_{,i}$ and 
$-\frac{5}{128}[\phi_{(2)}^3S_{(4),i}]_{,i}$. The result of the regularization 
method is stable against integration by parts only if the regularized value 
of the integral of any of these exact divergences is zero.

With $h_1$, Eq.\ (\ref{h1}), one can associate the following 18 exact 
divergences
\bea
\label{a11}
&\dst -\frac{5}{128}\left[\left(\phi_{(2)}^3\right)_{,i}S_{(4)}\right]_{,i},\,\,
-\frac{5}{128}\left[\phi_{(2)}^3S_{(4),i}\right]_{,i},\,\,
\frac{11}{32}
\left[\left(\phi_{(2)}\phi_{(4)}\right)_{,i}S_{(4)}\right]_{,i},\,\,
\frac{11}{32}\left[\phi_{(2)}\phi_{(4)}S_{(4),i}\right]_{,i},\,\,&
\nonumber\\[2ex]&\dst
-\frac{1}{8}\left[\phi_{(6)1,i}S_{(4)}\right]_{,i},\,\,
-\frac{1}{8}\left[\phi_{(6)1}S_{(4),i}\right]_{,i},\,\,
-\frac{13}{128}\left[\left(\phi_{(2)}^2\right)_{,i}S_{(6)}\right]_{,i},\,\,
-\frac{13}{128}\left[\phi_{(2)}^2S_{(6),i}\right]_{,i},\,\,&
\nonumber\\[2ex]&\dst
\frac{3}{16}\left[\phi_{(4),i}S_{(6)}\right]_{,i},\,\,
\frac{3}{16}\left[\phi_{(4)}S_{(6),i}\right]_{,i},\,\,
-\frac{7}{64}\left[\phi_{(2),i}S_{(8)}\right]_{,i},\,\,
-\frac{7}{64}\left[\phi_{(2)}S_{(8),i}\right]_{,i},\,\,&
\nonumber\\[2ex]&\dst
\frac{1}{4}\left[{\ghttiv ijk}S_{(6)ij}\right]_{,k},\,\,
\frac{1}{4}\left[{\httiv ij}S_{(6)ij,k}\right]_{,k},\,\,
\frac{3}{8}\left[\phi_{(2),i}\phi_{(4)}{\httiv ij}\right]_{,j},\,\,
\frac{3}{8}\left[\phi_{(2)}\phi_{(4),i}{\httiv ij}\right]_{,j},\,\,&
\nonumber\\[2ex]&\dst
-\frac{1}{16}\left[\phi_{(2),i}S_{(4)}{\httiv ij}\right]_{,j},\,\,
-\frac{1}{16}\left[\phi_{(2)}S_{(4),i}{\httiv ij}\right]_{,j}.&
\eea
The question now is whether, when explicitly computed (and regularized) by the 
method we are going to recall, the integrals of the exact divergences in the 
list (\ref{a11}) vanish or not. The explicit method of regularized-integration 
we use is the following: First, we explicitly perform the differentiations 
present in the divergences using, if necessary, the distributional rule of 
differentiation of homogeneous functions described in Appendix B 4 of 
\cite{JS98} (cf.\ the example given there).
Note that we always use Leibniz's rule when differentiating products of
functions, the distributional rule is used only for individual functions (by 
which we understand the functions explicitly entering Eqs.\ (\ref{a9}): 
$\phi_{(2)}$, $\phi_{(4)}$, $\phi_{(6)1}$, $S_{(4)}$, \ldots; it implies that in 
computing the new reference Hamiltonian given by Eqs.\ (\ref{a9}) we 
have not used the distributional rule).
E.g., we compute $\left(\phi_{(2)}^2\phi_{(2),j}\right)_{,i}$ in two steps:
(i) using Leibniz's rule one gets
$\left(\phi_{(2)}^2\phi_{(2),j}\right)_{,i}
=2\phi_{(2)}\phi_{(2),i}\phi_{(2),j}+\phi_{(2)}^2\phi_{(2),ij}$, (ii) the
distributional rule is applied only to $\phi_{(2),ij}$.
After this, a typical integral consists of two parts: without and with Dirac 
deltas. The first part is computed by means of the Riesz-formula-based 
regularization technique, while, for the second part, we have used the 
Hadamard-partie-finie regularization described in Appendix B 1 of \cite{JS98}. 
In the second case we have also investigated the stability of the result against 
`threading' the partie finie (Pf) over a product of functions, i.e., we have 
checked whether
\be
\label{a12}
{\rm Pf}(fg) = {\rm Pf}(f)\,{\rm Pf}(g).
\ee
For all divergences in (\ref{a11}) the final result of these explicit
computations is zero (regardless which 
side of Eq.\ (\ref{a12}) is employed in the contact terms).

With $h_2$, Eq.\ (\ref{h2}), one can similarly associate the following six 
integrals of exact divergences
\begin{mathletters}
\label{a13}
\bea
\Delta_{21} &\equiv& -\frac{1}{32} \int d^3\!x
\left[\phi_{(2)}^2\phi_{(6)1,i}\right]_{,i},
\\[2ex]
\Delta_{22} &\equiv& -\frac{1}{32} \int d^3\!x
\left[\left(\phi_{(2)}^2\right)_{,i}\phi_{(6)1}\right]_{,i},
\\[2ex]
\Delta_{23} &\equiv& -\frac{3}{32} \int d^3\!x
\left[\left(\phi_{(2)}\phi_{(4)}\right)_{,i}\phi_{(4)}\right]_{,i},
\\[2ex]
\Delta_{24} &\equiv& -\frac{3}{32} \int d^3\!x
\left[\phi_{(2)}\phi_{(4)}\phi_{(4),i}\right]_{,i},
\\[2ex]
\Delta_{25} &\equiv& -\frac{1}{16} \int d^3\!x
\left[\left(\phi_{(4)}^2\right)_{,i}\phi_{(2)}\right]_{,i},
\\[2ex]
\Delta_{26} &\equiv& -\frac{1}{16} \int d^3\!x
\left[\phi_{(4)}^2\phi_{(2),i}\right]_{,i}.
\eea
\end{mathletters}
Below we denote by primes the result of regularizing with the use of the 
left-hand side of Eq.\ (\ref{a12}) (for these parts of the integrands which are 
proportional to Dirac deltas), and by double primes the result obtained when
using the right-hand side of Eq.\ (\ref{a12}). 
With this notation, the regularized values of the integrals in 
(\ref{a13}) read
\begin{mathletters}
\label{a14}
\bea
&\dst \Delta_{21}' = \frac{1}{4}\frac{\nu}{r^4},\,\,\Delta_{21}'' = 0,&
\\[2ex]
&\Delta_{22}' = \Delta_{23}' = \Delta_{24}' = \Delta_{25}' = \Delta_{26}' = 0,&
\\[2ex]
&\Delta_{22}'' = \Delta_{23}'' = \Delta_{24}''
= \Delta_{25}'' = \Delta_{26}'' = 0.&
\eea
\end{mathletters}

With $h_3$, Eq.\ (\ref{h3}), we associate eight integrals of exact divergences
\begin{mathletters}
\label{a15}
\bea
\Delta_{31} &\equiv& 2 \int d^3\!x
\left[\left(2\pi_{(3)}^i+\Delta^{-1}\pi_{(3),il}^l\right)_{,k}
{\pitiii jk}{\httiv ij}\right]_{,k},
\\[2ex]
\Delta_{32} &\equiv& -\int d^3\!x
\left[\left(2\pi_{(3)}^i+\Delta^{-1}\pi_{(3),il}^l\right)
\left(2\pi_{(3)}^k+\Delta^{-1}\pi_{(3),km}^m\right){\ghttiv ijk}\right]_{,j},
\\[2ex]
\Delta_{33} &\equiv& -\int d^3\!x
\left[\left(2\pi_{(3)}^i+\Delta^{-1}\pi_{(3),il}^l\right)
\left(2\pi_{(3)}^k+\Delta^{-1}\pi_{(3),km}^m\right)_{,j}{\httiv ij}\right]_{,k},
\\[2ex]
\Delta_{34} &\equiv& -\frac{5}{128}\int d^3\!x 
\left[\phi_{(2)}^2\phi_{(2),j}{\httiv ij}\right]_{,i},
\\[2ex]
\Delta_{35} &\equiv& \frac{1}{4}\int d^3\!x
\left[{\httiv ij}\left(\phi_{(2)}{\httiv ij}\right)_{,k}\right]_{,k},
\\[2ex]
\Delta_{36} &\equiv& \frac{1}{4}\int d^3\!x
\left[\phi_{(2)}{\httiv ij}{\ghttiv ijk}\right]_{,k},
\\[2ex]
\Delta_{37} &\equiv& \frac{5}{8} \int d^3\!x
\left[\left(\phi_{(2)}{\httiv ij}\right)_{,k}S_{(4)ij}\right]_{,k},
\\[2ex]
\Delta_{38} &\equiv& \frac{5}{8} \int d^3\!x
\left[\phi_{(2)}{\httiv ij}S_{(4)ij,k}\right]_{,k}.
\eea
\end{mathletters}
The results of the regularized-integration of (\ref{a15}) read (here 
$\kappa\equiv\nu^2\left(\pp-3\np^2\right)/r^3$; note that the exact divergences 
$\Delta_{32}$ and $\Delta_{34}$ do not contain contact terms)
\begin{mathletters}
\label{a16}
\bea
&\dst \Delta_{31}' = \frac{7}{10}\kappa,\quad
\Delta_{31}'' = -\frac{109}{10}\kappa,&
\\[2ex]
&\dst \Delta_{32} = -\frac{11}{30}\kappa,&
\\[2ex]
&\dst \Delta_{33}' = -\frac{21}{10}\kappa,\quad
\Delta_{33}'' = \frac{37}{20}\kappa,&
\\[2ex]
&\dst \Delta_{34} = -\frac{1}{2}\kappa,&
\\[2ex]
&\dst \Delta_{35}' = \frac{32}{25}\kappa,\quad
\Delta_{35}'' = -\frac{21}{10}\kappa,&
\\[2ex]
&\dst \Delta_{36}' = \frac{32}{25}\kappa,\quad
\Delta_{36}'' = 0,&
\\[2ex]
&\dst \Delta_{37}' = -8\kappa,\quad
\Delta_{37}'' = 0,&
\\[2ex]
&\dst \Delta_{38}' = 0,\quad
\Delta_{38}'' = 8\kappa.&
\eea
\end{mathletters}

{From} Eqs.\ (\ref{a14}) and (\ref{a16}) follows that only one out of the 32
exact divergences considered above, $\Delta_{21}=\nu/(4r^4)$, is connected to
the static ambiguity parameter $\oms$, whereas eight exact divergences
$\Delta_{31}$--$\Delta_{38}$ `contribute' to the kinetic ambiguity parameter
$\omk$; their regularized values are between $-10.9\kappa$ and $+8\kappa$.

We are not confident that the so-found relatively small contribution ($ \pm
1/4$) to $\oms$ is indicative of a smaller ambiguity in $\oms$ than in $\omk$.
Therefore, we used also another (more primitive, but hopefully more robust) way
of estimating the plausible values of the ambiguity parameters $\oms$ and
$\omk$.  We simply looked at the values of all the numerical coefficients in the
final 3PN Hamiltonian, Eq.\ (\ref{36}).  The Hamiltonian has 27 numerical
coefficients, entering as factors of the different monomials made of powers of
$\pp$, $\np^2$, $1/r$, and $\nu$.  Out of them only three are influenced by
ambiguities (the coefficients of $\nu^2\pp/r^3$, $\nu^2\np^2/r^3$, and
$\nu/r^4$).  The numerical values of the unambiguous coefficients range between
$\pi^2/64-335/48\approx-6.825$ ($\nu\,\pp/r^3$) and $17/2=8.5$
($\nu\,\ppp^2/r^2$), whereas the ambiguous coefficients in the reference
Hamiltonian (\ref{36}) are equal to $-55/12\approx-4.583$ ($\nu^2\pp/r^3$),
$27/8=3.375$ ($\nu^2\np^2/r^3$), and $109/12-21\pi^2/32\approx2.606$
($\nu/r^4$).

As a result of the above discussion we take, as plausible ranges for the 
ambiguity parameters $\oms$ and $\omk$, the ranges
\be
-10 \lesssim \oms \lesssim 10,\quad -10 \lesssim \omk \lesssim 10.
\ee

\section{Computation of the radial action integral}

The explicit form of the coefficients $A$, $B$, $C$, $D_i$ entering the 
effective radial potential $R(r,E,j)$, Eq.\ (\ref{44}), reads
\begin{mathletters}
\bea
A &=& 2E + \frac{1}{c^2} (1-3\nu)\,E^2 - \frac{1}{c^4} (1-4\nu)\nu\,E^3
+  \frac{1}{c^6} \frac{5}{4}(1-4\nu)\nu^2\,E^4,
\\[2ex]
B &=& 1 + \frac{1}{c^2} (4-\nu)\,E + \frac{1}{c^4} (2-2\nu+\nu^2)\,E^2
+  \frac{1}{c^6} (2-\nu)\nu^2\,E^3,
\\[2ex]
C &=& -j^2 + \frac{1}{c^2} (6+\nu) + \frac{1}{c^4} 15E
+  \frac{1}{c^6} \frac{1}{6}(45-38\nu-4\nu^2)\,E^2,
\\[2ex]
D_1^{\text{1PN}} &=& -\frac{1}{c^2} \nu\,j^2
+ \frac{1}{c^4} \left[ \frac{1}{2}(17+5\nu+4\nu^2)-(1+\nu)\nu\,Ej^2 \right]
\nonumber\\[2ex]&&
+ \frac{1}{c^6} \left\{
\frac{1}{8}\left[144+(\pi^2-212)\nu+16(4\omk-7)\nu^2+16\nu^3\right]E
- \frac{1}{2}\nu^2\,E^2j^2 \right\},
\\[2ex]
D_2^{\text{2PN}} &=& -\frac{1}{c^4} (1+3\nu)\nu\,j^2
+ \frac{1}{c^6} \left\{ \frac{1}{48}
\left[384+(69\pi^2-1660-96\oms)\nu+4(96\omk-139)\nu^2+240\nu^3\right]
\right.\nonumber\\[2ex]&&\left.
+ \frac{1}{12}(79+38\nu-72\nu^2)\nu\,Ej^2 \right\},
\\[2ex]
D_3^{\text{2PN}} &=& \frac{1}{c^4} \frac{3}{4}\nu^2\,j^2
+ \frac{1}{c^6} \left\{
\frac{1}{96}\left[140-9\pi^2 +8(191-72\omk)\nu-960\nu^2\right]\nu\,j^2
+ \frac{9}{4}\nu^3\,Ej^4 \right\},
\\[2ex]
D_4^{\text{3PN}} &=& -\frac{1}{c^6}
\frac{1}{6}(5+28\nu-30\nu^2)\nu\,j^4,
\\[2ex]
D_5^{\text{3PN}} &=& -\frac{1}{c^6} \frac{5}{8}\nu^3\,j^6.
\eea
\end{mathletters}

The PN expansion of the effective radial action integral $i_r$, Eq.\ (\ref{45}), 
can be written (as explained in Sec.\ IV of the present paper) as a linear 
combination of the integrals $J_{q,m}$ defined in Eq.\ (\ref{410}).
The only integrals that one needs to compute at 3PN order are $J_{q,m}$ with 
$0\le q\le3$: when $q=0$, $m=0$; when $q=1$, $3\le m\le7$; when $q=2$, $6\le 
m\le8$; and when $q=3$, $m=9$. The explicit results of the integration
(performed by the two methods indicated in the text) read
\begin{mathletters}
\bea
J_{0,0} &=& \frac{B}{\sqrt{-A}}-\sqrt{-C},
\\[2ex]
J_{1,3} &=& \frac{B}{(-C)^{\frac{3}{2}}},
\\[2ex]
J_{1,4} &=& \frac{3B^2-AC}{2(-C)^{\frac{5}{2}}},
\\[2ex]
J_{1,5} &=& \frac{B(5B^2-3AC)}{2(-C)^{\frac{7}{2}}},
\\[2ex]
J_{1,6} &=& \frac{35B^4-30AB^2C+3A^2C^2}{8(-C)^{\frac{9}{2}}},
\\[2ex]
J_{1,7} &=& \frac{B(63B^4-70AB^2C+15A^2C^2)}{8(-C)^{\frac{11}{2}}},
\\[2ex]
J_{2,6} &=& \frac{3(-5B^2+AC)}{2(-C)^{\frac{7}{2}}},
\\[2ex]
J_{2,7} &=& \frac{5B(-7B^2+3AC)}{2(-C)^{\frac{9}{2}}},
\\[2ex]
J_{2,8} &=& \frac{15(-21B^4+14AB^2C-A^2C^2)}{8(-C)^{\frac{11}{2}}},
\\[2ex]
J_{3,9} &=& \frac{35B(3B^2-AC)}{2(-C)^{\frac{11}{2}}}.
\eea
\end{mathletters}

\end{document}